\documentclass[10pt,prx,aps,twocolumn,showpacs,floatfix,superscriptaddress]{revtex4-1}
\usepackage{graphicx}
\usepackage{amsmath}
\usepackage{bm} 
\usepackage{color}

\begin{document}


\title{Unifying static and dynamic properties in 3D quantum antiferromagnets}

\author{H. D. Scammell}
\affiliation{School of Physics, The University of New South Wales,
  Sydney, NSW 2052, Australia}
\author{Y. Kharkov}
\affiliation{School of Physics, The University of New South Wales,
  Sydney, NSW 2052, Australia}
\author{Yan Qi Qin}
\affiliation{Beijing National Laboratory for Condensed Matter Physics and 
  Institute of Physics, Chinese Academy of Sciences, Beijing 100190, China}
\author{Zi Yang Meng}
\affiliation{Beijing National Laboratory for Condensed Matter Physics and 
  Institute of Physics, Chinese Academy of Sciences, Beijing 100190, China}
\author{B. Normand}
\affiliation{Neutrons and Muons Research Division, Paul Scherrer Institute, 
  CH-5232 Villigen-PSI, Switzerland}
\author{O. P. Sushkov}
\affiliation{School of Physics, The University of New South Wales,
  Sydney, NSW 2052, Australia}

\date{\today}

\begin{abstract}

Quantum Monte Carlo simulations offer an unbiased means to study the 
static and dynamic properties of quantum critical systems, while quantum 
field theory provides direct analytical results. We study three dimensional, 
critical quantum antiferromagnets by performing a combined analysis using 
both quantum field theory calculations and quantum Monte Carlo data. 
Explicitly, we analyze the order parameter (staggered magnetization), 
N\'eel temperature, quasiparticle gaps, and the susceptibilities in the 
scalar and vector channels. We connect the two approaches by deriving 
descriptions of the quantum Monte Carlo observables in terms of the 
quasiparticle excitations of the field theory. The remarkable agreement 
not only unifies the description of the static and dynamic properties of 
the system, but also constitutes a thorough test of perturbative O(3) 
quantum field theory and opens new avenues for the analytical guidance 
of detailed numerical studies.

\end{abstract}

\maketitle

\section{Introduction}\label{Introduction}

Quantum field theories (QFTs) are of fundamental importance to both high-energy 
and statistical physics. In particular, the generic O($N$)-symmetric, 
$d$-dimensional field theory finds a remarkably broad application. For 
$N = 0$ this theory describes the self-avoiding random-walk problem, while 
for $N = 1$, $2$, and $3$ it describes magnetic models with, respectively, 
Ising, XY, and Heisenberg interactions. In nuclear physics, the $N = 4$ 
version in $d = 4$ dimensions is of particular importance because it provides 
an effective theory for $\pi$-mesons. Taking $N \rightarrow \infty$, one 
obtains the spherical model \cite{Zinn-Justin}. 

In the vicinity of a classical or a quantum phase transition (QPT), any 
characteristic length scale of a physical system diverges \cite{Sachdev2011}. 
If the system is described by a QFT, its properties then depend solely on the 
dimensionality, $d$, and the internal symmetries, which for O($N$) theories 
means the number of components, $N$. These provide a unique determination of 
the universality class and hence of the critical exponents of the field theory 
at the QPT. The robust predictions of QFT in this regard have inspired a 
multitude of experimental and numerical studies, and in fact constitutes an 
entire subfield of physics. 

Quite generally, quantum systems in high dimensions have sufficiently many 
degrees of freedom that their behavior is ``free,'' governed by the same set 
of exponents that can be derived at the mean-field level. Systems in low 
dimensions are constrained and their exponents are ``anomalous,'' depending 
in detail on $d$, $N$, and the form of the interaction terms. A situation 
of special importance occurs for systems at the upper critical dimension, 
$d_c = 4$, which in the quantum case is often expressed as $3+1$ [for three 
spatial and one temporal dimension(s)]. Here the critical exponents are 
predicted to take mean-field values, which for O($N$) field theories are 
independent of $N$, augmented by multiplicative logarithmic corrections to 
the observables. Because an explicit $N$-dependence does appear in the 
multiplicative logarithmic corrections, these represent a fundamental test 
of universality \cite{Zinn-Justin, Kenna1993, Kenna1994erratum} and their 
existence has profound consequences in both high-energy and statistical 
physics. 


Although there exists a wealth of analytical results detailing the theory 
of logarithmic corrections \cite{Zinn-Justin, Wegner1973, Domb1976, Hara1987, 
Fernandez1992, Li1997, Kleinert2001}, discerning them in experimental 
measurements is a hugely demanding task requiring datasets spanning many 
orders of magnitude in parameter space near a QPT. Similarly, their numerical 
determination in lattice simulations is a delicate and highly computationally 
intensive proposition. Numerical tests of logarithmic corrections have mostly 
been restricted to the $N = 1$ theory \cite{Kenna1993, Kenna1994, 
Kenna1994erratum, deForcrand2010}, and only recently has a movement beyond 
$N = 1$ been driven by a confluence of refined numerical methods, increasing 
computer power, and rising interest from experiments in condensed matter 
\cite{SachdevSolvay}. Experimental studies of QPTs were motivated initially 
by problems in superconductivity, where the order parameter has U(1) or 
equivalently O(2) symmetry, and have since broadened to include quantum 
magnetism, where the order parameter in the Heisenberg case has O(3) 
symmetry \cite{Chakravarty1988}, and condensates of ultracold atoms, in 
which different symmetries can be realized. In all cases the system 
dimensionality is $d = 1$, 2, or 3.

Here we specialize to the case of quantum antiferromagnets (QAFs). Critical 
magnetic systems in the $d = 2+1$, $N = 3$ universality class have been the 
object of extensive numerical \cite{Sandvik1994, Troyer1996, Matsumoto2001, 
Wang2006, Wenzel2009} and analytical \cite{Podolsky2011, Podolsky2012, 
Gazit2013, GazitArovas2013, Rancon2014} investigation for over two decades, 
and have undergone a recent revival due to their close parallels in ultracold 
atomic experiments. However, our present focus is the $d = 3+1$, $N = 3$ QPT, 
which on the theoretical side encompasses all the physics of the upper 
critical dimension and on the experimental side is realized in the compound 
TlCuCl$_3$. TlCuCl$_3$ is a $S = 1/2$ QAF with a dimerized geometry and 
three-dimensional (3D) interdimer coupling, which can be driven by an applied 
hydrostatic pressure through a QPT between a magnetically ordered AF phase and 
a ``quantum disordered'' dimerized phase. Elastic and inelastic neutron 
scattering experiments on TlCuCl$_3$ \cite{Ruegg2004, Ruegg2008, Merchant2014} 
have characterized clearly the hallmarks of the magnetic QPT in both the 
static and dynamic properties. 

\begin{figure}[t]
\hspace{0.05cm}{\includegraphics[width=0.415\textwidth,clip]{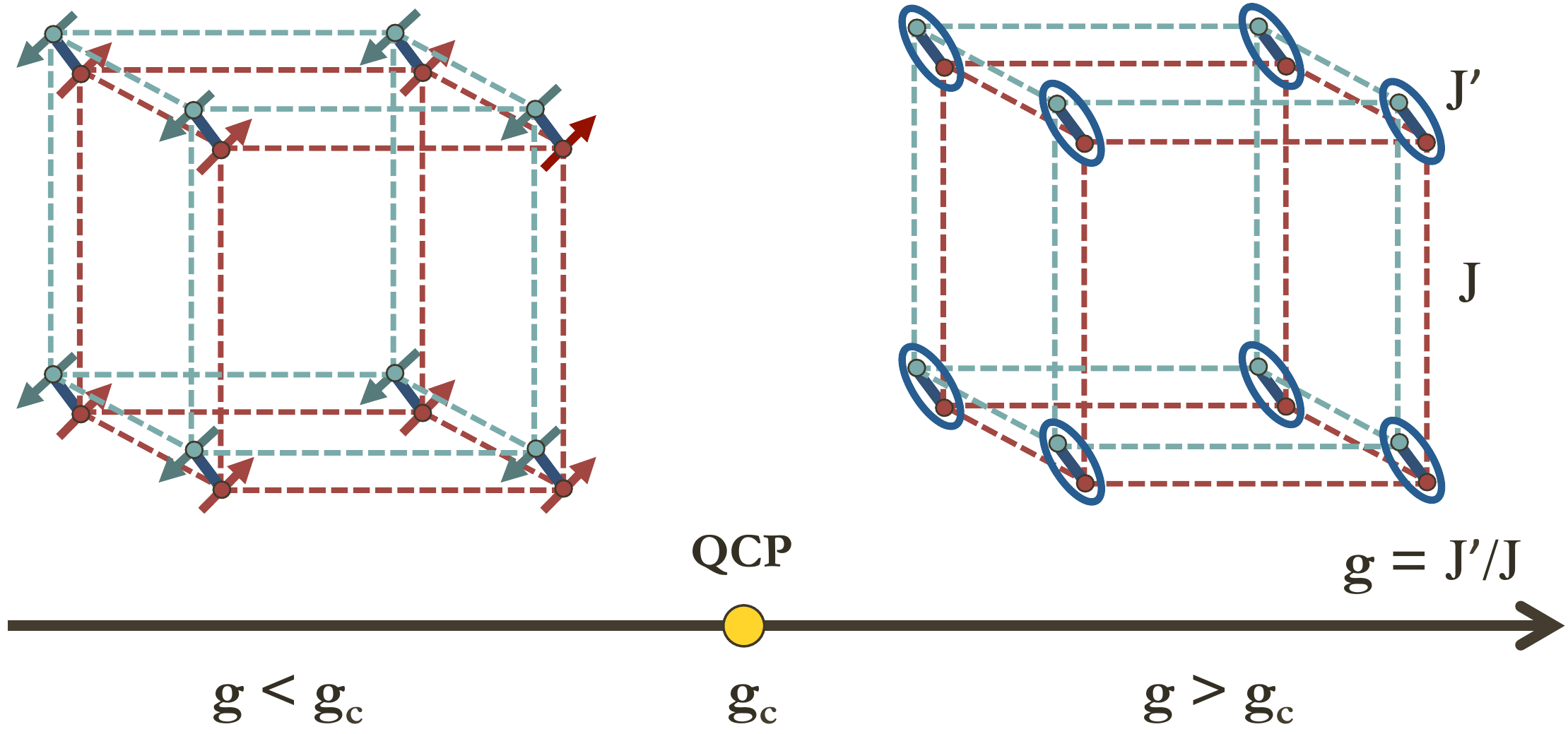}}
\hspace{0.7cm}{\includegraphics[width=0.31\textwidth,clip]{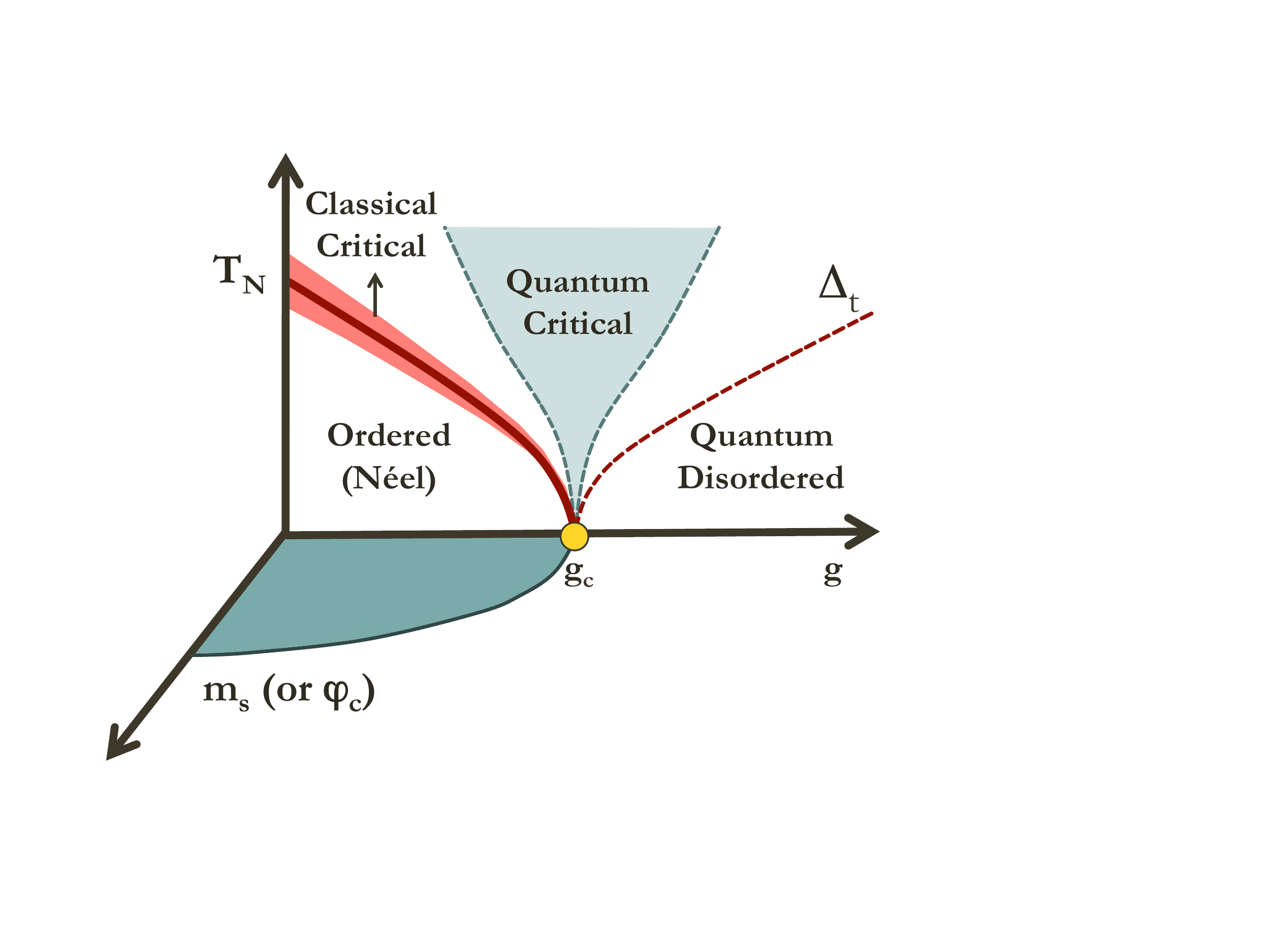}} 
\hspace{1.1cm}
\put(-200,157){\text{{\bf (a)}}} 
\put(-200,12){\text{{\bf (b)}}} 
\caption{(a) Dimerized lattice of $S = 1/2$ spins in the 3D double-cubic 
geometry. Sites of the red and blue cubic lattices are connected pairwise 
by dimer bonds. $J'$ and $J$ are antiferromagnetic Heisenberg interactions 
respectively on and between the dimer units. Their ratio, $g = J'/J$, 
controls the QPT from a N\'eel ordered phase (left) to a quantum disordered 
dimer-singlet phase (right). The QCP occurs at the critical ratio $g_c$. 
(b) Schematic quantum critical phase diagram for the Heisenberg model on 
the double-cubic lattice. The staggered magnetization, $m_s$ (or $\varphi_c$), 
N\'eel temperature, $T_N$, and triplon gap, $\Delta_t$, all vanish at the QCP. 
Not shown is the Higgs gap, $\Delta_H$, which is proportional to $\varphi_c$ 
and hence also vanishes at the QCP. }
\label{phase}
\end{figure}

From the viewpoint of QFT, the 3D dimerized QAF provides an excellent test 
case for the study of critical properties around the QPT at $d_c$. The 
weakness of QFTs is that, as effective low-energy, long-wavelength theories, 
their connection to real systems is only through phenomenological parameters, 
and thus it is essential to benchmark them against numerical and experimental 
realizations. Indeed the effective O(3), $d = 3+1$ QFT has already been used 
to provide an accurate analytical description of the critical properties 
observed in TlCuCl$_3$ \cite{SachdevSolvay, Kulik2011, ScammellFreedom2015, 
Katan2015, ScammellWidths2017, Fidrysiak2017}. Numerically, the method of 
choice for computing the properties of the unfrustrated QAF is Quantum Monte 
Carlo (QMC), with which recent large-scale simulations of the 3D dimerized 
QAF across the quantum critical regime have been performed for $S = 1/2$ 
spins with Heisenberg interactions on the double-cubic geometry depicted 
in Fig.~\ref{phase}(a). First a systematic study of the static properties 
by some of us \cite{Qin2015} demonstrated to high precision the validity of 
the theoretical predictions concerning multiplicative logarithmic corrections 
for this universality class. Next, two parallel studies \cite{Qin2017, 
Lohofer2017} used QMC and analytic continuation methods to access the 
dynamical properties of the system. The aim of the present work is, within 
a one-loop perturbative renormalization-group (RG) treatment of the O(3), 
$d = 3+1$ QFT, to analyze and unify the static and dynamic observables 
obtained in these QMC simulations.

In the vicinity of the magnetic quantum critical point (QCP), the observables 
accounting for the relevant (critical) degrees of freedom are associated with 
the broken or unbroken O(3) symmetry. In the symmetric (quantum disordered) 
phase there are three degenerate, gapped spin excitations, which because of 
their triplet character are known as triplons; their energy gap, denoted by 
$\Delta$ in Fig.~\ref{phase}(b), closes as the QCP is approached. In the 
symmetry-broken phase, a preferred direction is established and is associated 
with an order parameter, which for a QAF is the staggered magnetization, $m_s$. 
In three spatial dimensions, magnetic order is present up to a finite N\'eel 
temperature, $T_N$, at which it is destroyed by thermal fluctuations. An 
illustration of the phase diagram and the behavior of these observables is 
presented in Fig.~\ref{phase}(b). 

Directional oscillations of the order parameter are acoustic (gapless) and 
are are known as Goldstone modes. Their linear dispersion about the gapless 
point ensures that the dynamical critical exponent is $z = 1$, and hence 
that the time axis counts as one additional system dimersion. By contrast, 
the amplitude oscillation of the order parameter is a gapped mode, often 
referred to as the ``Higgs mode'' by analogy with the amplitude modes in 
a superconductor and in electroweak field theory, although in the QAF it 
lacks the gauge character of these two systems. In the O(3) case there are 
two Goldstone modes and one Higgs, such that the three modes of the phases 
on either side of the QPT evolve continuously evolve into each other at the 
QCP. Because of its finite gap, or mass, it is possible in the O(3) QFT for 
the Higgs mode to decay spontaneously into Goldstone modes, and therefore 
it has not only an energy but also an intrinsic line width. 

QFT and QMC both provide direct access to the static quantities of the system, 
namely the staggered magnetization and N\'eel temperature, and to the dynamic 
ones, which are the characteristic energy gaps of the triplon and Higgs modes, 
as well as the Higgs decay width. In QMC, the static and dynamic quantities 
are treated on a quite unequal footing, requiring very different techniques 
to extract. By contrast, they appear in a completely symmetric way in a QFT 
and thus are treated on an equal footing, being equivalently and uniquely 
determined by a set of (five) phenomenological QFT parameters. However, where 
a QFT is an effective low-energy theory, the applicability of QMC is by no 
means limited to the low-energy sector, nor by any of the other approximations 
inherent to QFT, and in this sense QMC is a completely unbiased method.

The static \cite{Qin2015} and dynamic \cite{Qin2017, Lohofer2017} observables 
computed by QMC on both sides of the QCP for the double-cubic QAF have each 
been shown to fit the universal scaling forms expected from the O(3) QFT 
with $d = 3+1$ \cite{Zinn-Justin}, including their logarithmic corrections. 
Nevertheless, important questions remain for both QMC and QFT. Specifically, 
space-time symmetry is largely lost in QMC, and with it any underlying 
connection between static and dynamic variables. While QFT is in principle 
perfectly suited for retrieving this connection, it has yet to be determined 
whether or not all of the observables of the system can be described 
quantitatively by an effective low-energy QFT with a single set of 
phenomenological parameters. An alternative statement of our primary goal 
is to derive this single set of parameters. 

Further, the Higgs line width is an important additional observable but its 
determination lies at the limits of current numerical capabilities. The 
vector and scalar response functions used to compute the Higgs decay rate 
in the recent QMC studies \cite{Qin2017, Lohofer2017} are described naturally 
by QFT in terms of the Green functions, or generalized response functions, of 
the magnetic excitations (the Goldstone and Higgs modes). Thus one may perform 
a detailed analysis of the vector and scalar response functions to obtain 
analytical guidance for interpreting the existing QMC line-width data and 
for structuring future numerical studies.  

This paper is organized as follows. In Sec.~\ref{Model} we present the 
lattice Hamiltonian we study, summarize the QMC methods we have applied and 
the nature of their output, formulate the QFT description at the mean-field 
level, and detail the process for computing one-loop RG corrections. In 
Sec.~\ref{Results} we apply the analytical QFT formulas to fit the static 
and dynamic QMC data of Refs.~\cite{Qin2015, Qin2017} and extract the 
phenomenological QFT parameters. Section \ref{Decay} provides a detailed 
analysis of the vector and scalar response functions, with which we analyze 
the Higgs line width for comparison with QMC \cite{Qin2017}. For completeness, 
in Sec.~\ref{derivation} we relate the optimal QFT parameters to the analogous 
quantities derived from a microscopic description, for which we use a 
bond-operator framework. In Sec.~\ref{discussion} we discuss the context 
of our results and their value for future research directions.

\section{Model and Methods}\label{Model}

The double-cubic geometry, shown in Fig.~\ref{phase}(a), is perhaps the 
most representative and spatially symmetric 3D dimerized lattice. This 
system consists of two interpenetrating simple cubic lattices with the same 
antiferromagnetic interaction strength, $J$, connected pairwise by another 
antiferromagnetic interaction, $J'$; there is no frustration in this 
situation. The ground state for low coupling ratios, $g = J'/J$, is a 
N\'eel-ordered phase of finite staggered magnetization  and for high $g$ it 
is dimer-singlet phase with no order, as illustrated in Fig.~\ref{phase}(b). 
The critical coupling ratio for the QPT is denoted by $g_c$. The Hamiltonian 
is 
\begin{align}
\label{Hamiltonian}
H = J\sum_{<i,j>} \{\bm S_l^i \cdot \bm S_l^j + \bm S_r^i \cdot \bm S_r^j \}
 + J'\sum_{i} \bm S_l^i\cdot \bm S_r^i,
\end{align}
where the subscripts $l$ and $r$ denote the two spins on a single dimer bond. 

Supplemented by an appropriate treatment of the temperature, 
Eq.~(\ref{Hamiltonian}) contains all of the information about the static 
and dynamic observables of the system, whose qualitative behavior is depicted 
in Fig.~\ref{phase}(b). We summarize the two parallel techniques we employ 
here to extract those observables, namely direct numerical QMC simulation 
augmented by stochastic analytic continuation (SAC), and the analysis of the 
effective low-energy QFT derived from Eq.~(\ref{Hamiltonian}). Both QMC and 
QFT techniques allow for the efficient inclusion of finite temperatures 
in the quantum system, albeit in very different ways we outline below.  

\subsection{Quantum Monte Carlo}

We have performed QMC simulations using the stochastic series expansion (SSE) 
technique \cite{SandvikKurkijarvi1991, SandvikRapid1999, Evertz2003}. In this 
method, spin configurations are constructed in the $S^z$ basis, evolved over 
an imaginary time $\tau$, and sampled systematically. The (squared) order 
parameter is evaluated straightforwardly from the spatial and temporal 
average of $S_i^{lz}(\tau) - S_i^{rz}(\tau)$ and dynamical correlation functions 
are obtained from operator strings connecting states $S_i^{lz}(\tau_1) - S_i^{rz}
(\tau_1)$ and $S_j^{lz}(\tau_2) - S_j^{rz}(\tau_2)$. To avoid the repetition of 
published material, we refer the reader to Ref.~\cite{Qin2015}. To evaluate 
the static quantities, we have performed simulations on cubic systems of 
$2L^3$ sites for values of $L$ up to and including 48, and at temperatures 
down to $1/2L$. By detailed finite-size scaling we extrapolate to the 
thermodynamic limit to obtain unbiased results with well-controlled 
statistical errors. We comment that the errors on $m_s$ in the ordered 
phase, which extrapolates to a finite zero-temperature quantity for all 
$g < g_c$, are significantly smaller than the errors on $T_N$, which are 
determined from the vanishing of $m_s$ at finite temperatures. 

In order to obtain the dynamical response of the system, we first measure 
the imaginary-time structure factor and then employ SAC \cite{Sandvik1998, 
Beach2004, Syljuasen2008, Fuchs2010, Sandvik2016} to obtain the real-frequency 
spectral function. This process can be performed using both the spin operator, 
$S_i^{lz}(\tau) - S_i^{rz}(\tau)$, and the dimer operator, $B_i(\tau) = \bm 
S_i^l(\tau) \cdot \bm S_i^r(\tau) - \langle \bm S_i^l(\tau) \cdot \bm S_i^r(\tau) 
\rangle$. The spin spectral function is referred to as the vector response 
function or the $S = 1$ channel and the dimer spectral function as the scalar 
response function or $S = 0$ channel. Again we refer the reader to previously 
published material \cite{Qin2017}. Because the extraction of dynamical 
quantities is considerably more computationally intensive, our maximum $L$ is 
limited to 24 and the errors in extrapolated quantities are correspondingly 
larger, but still well characterized. In these simulations the excitation 
gaps, $\Delta_t$ for the triplons at $g > g_c$ and $\Delta_H$ for the Higgs 
mode at $g < g_c$, are obtained with significantly greater accuracy than the 
Higgs line width, $\Gamma_H$, obtained from either channel. We note that 
the present study does not involve any new simulations, but that we have 
reanalyzed some of our existing data \cite{Qin2015, Qin2017} in the light 
of the comparison with QFT. 

At zero temperature and in the quantum critical regime, the observables 
$m_s$, $\Delta_t$, and $\Delta_H$ have the generic form of a power-law 
dependence on the separation from the QCP, $\delta g = (g - g_c)/g_c$, 
multiplied by a logarithmic correction \cite{Zinn-Justin,ScammellFreedom2015, 
Qin2015,Qin2017,Lohofer2017}. We express them in the form 
\begin{align}
\label{observablesms}
m_s(g) & = a_1 |g - g_c|^{\nu_1} \ln \left[ \frac{|g - g_c|}{b_1} 
\right]^{\beta_1}, \\
\label{observablest}
\Delta_t(g) & = a_2 |g - g_c|^{\nu_2} \ln \left[ \frac{|g - g_c|}{b_2} 
\right]^{\beta_2}, \\
\label{observablesH}
\Delta_H(g) & = a_3 |g - g_c|^{\nu_3} \ln \left[ \frac{|g - g_c|}{b_3} 
\right]^{\beta_3}.
\end{align}
At finite temperatures, the N\'eel temperature can be expressed in the same 
manner \cite{Qin2015, ScammellFreedom2015}, as 
\begin{align}
\label{observablesTN}
T_N(g) & = a_4 |g - g_c|^{\nu_4} \ln \left[ \frac{|g - g_c|}{b_4} \right]^{\beta_4}.
\end{align}
The quantum critical behavior is then gathered in the exponents $\nu_i$ for 
the power-law dependence and $\beta_i$ for the multiplicative logarithmic 
correction. The exponents $\{\nu_i, \beta_i\}$ have received a great deal of 
attention and have been discussed by scaling hypotheses and general QFT  
arguments for many different universality classes. At the upper critical 
dimension, $\nu_i = 1/2$, i.e.~all observables follow a predominantly 
mean-field form, independent of $N$. For an O($N$) system at $d_c$, the 
static observables have $\beta_1 = \beta_4 = 3/(N+8)$ at one-loop order 
and the dynamic observables have $\beta_2 = \beta_3 = - (N+2)/2(N+8)$ 
\cite{Zinn-Justin}. Although these critical exponents have been verified to 
high precision by the recent QMC analyses \cite{Qin2015, Qin2017, Lohofer2017}, 
the relationships among the coefficients $\{a_i, b_i\}$ remains unknown and 
can be determined by appealing to QFT.

\subsection{Quantum field theory: Mean-field treatment}\label{QFT}

To capture the ordered and disordered phases, the QPT between them, 
and the low-energy magnetic degrees of freedom, we adopt the effective 
description of the Hamiltonian (\ref{Hamiltonian}) provided by the 
Lagrangian field theory \cite{Sachdev2011} 
\begin{align}
\label{Lagrangian}
{\cal L} & = {\textstyle \frac{1}{2}} \partial_{\mu} {\vec{\varphi}} \, 
\partial^{\mu} {\vec{\varphi}} - {\textstyle \frac{1}{2}} m^2 
{\vec{\varphi}}^{\ 2} - {\textstyle \frac{1}{4}} \alpha [\vec{\varphi}^{\ 2}]^{2}.
\end{align}
Here $\vec{\varphi}$ is a vector field describing the staggered magnetization, 
$m$ is a mass term for free field fluctuations, $\alpha$ is a stiffness term 
governing the interactions of $\vec{\varphi}$ fluctuations, and the index 
$\mu$ enumerates one time and three space coordinates, with $\partial_{\mu} = 
(\partial_t, c\nabla)$, where the constant of proportionality, $c$, is the 
velocity of the Goldstone modes in the ordered phase. For later quantitative 
purposes (Secs.~\ref{QFToneloop} and \ref{Results}A) we note that $m$ is 
defined to have units of energy (and $\alpha$ of an energy cubed). 

Qualitatively, the QPT is controlled in Eq.~(\ref{Lagrangian}) through the 
mass term, which we express at linear order as $m^2(\delta g) = \gamma^2 
(g - g_c)/g_c$, where $\gamma^2 > 0$ is another constant of proportionality. 
For $g > g_c$, $m^2 > 0$ and the classical expectation value of the field is 
$\varphi_c^2 = 0$, which describes the magnetically disordered phase. The 
system has a global rotational symmetry and its excitations (the triplons) 
are gapped and triply degenerate. For $g < g_c$, $m^2 < 0$ and the (staggered) 
field takes a non-zero classical expectation value, $\varphi^2_c = |m^2|/ 
\alpha$, which describes the ordered antiferromagnetic phase. Changing $m^2$ 
from positive to negative causes a spontaneous breaking of the O(3) spin 
symmetry and the excitations of the symmetry-broken phase are two gapless, 
transverse excitations (spin waves, the Goldstone modes) and one gapped, 
longitudinal excitation (the amplitude or Higgs mode). It is straightforward 
using the bare (unrenormalized) parameters to note that the triplon gap (at 
$g > g_c$) is $\Delta_t(\delta g) = m(\delta g)$ and the Higgs gap (at $g < 
g_c$) is $\Delta_H(\delta g) = \sqrt{2} |m(\delta g)|$, and hence to recover 
the relation $\Delta_H/\Delta_t = \sqrt{2}$.

\begin{figure}[t]
{\hspace{0.75cm}\includegraphics[width=0.35\textwidth,clip]{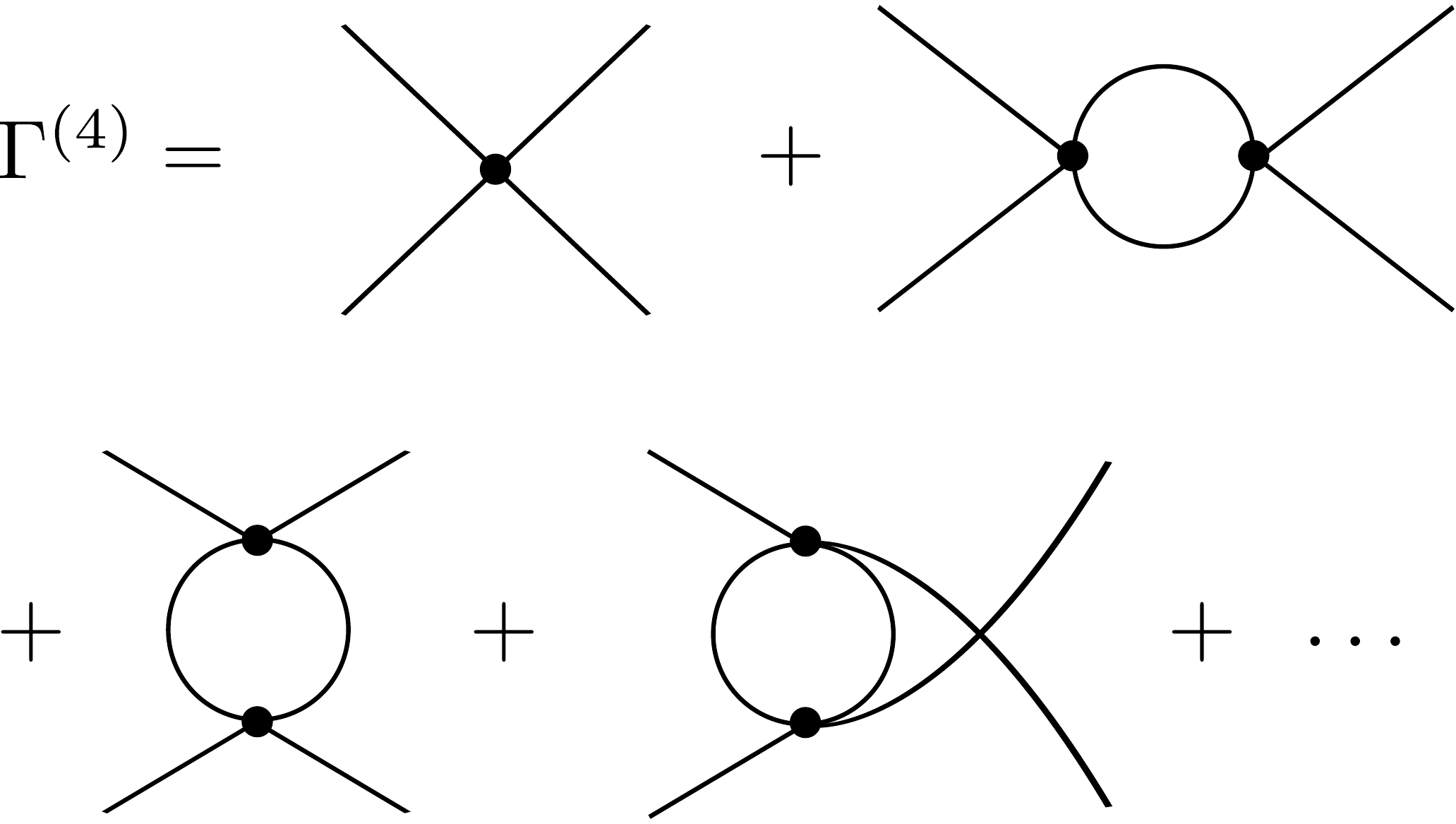}}
\vspace{0.5cm} \\
{\hspace{0.75cm}\includegraphics[width=0.35\textwidth,clip]{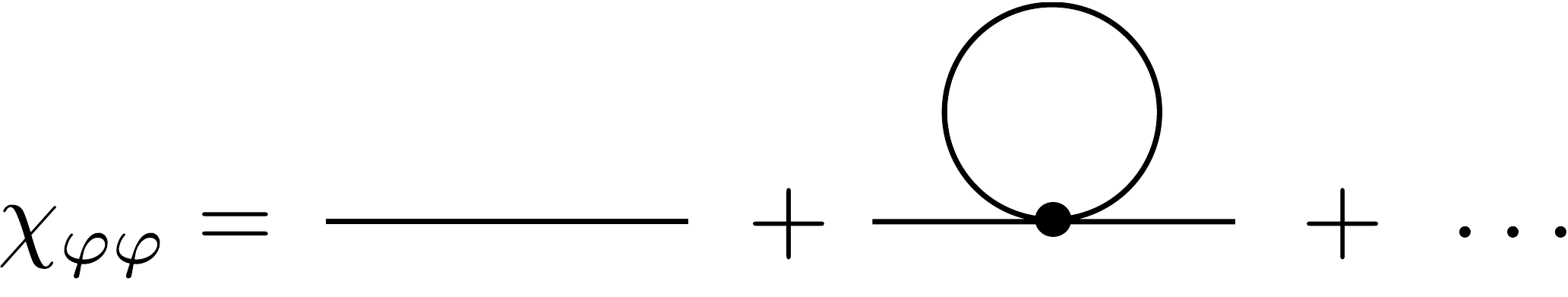}}
\begin{picture}(0,0) 
\put(-210,100){\text{{\bf (a)}}} 
\put(-210,10){\text{{\bf (b)}}} 
\end{picture}
\caption{Diagrammatic expansions for (a) the four-point vertex, $\Gamma^{(4)}$,
and (b) the response function, $\chi_{\varphi\varphi}$, shown for the 
quantum disordered phase ($g > g_c$). Solid lines denote the free propagation, 
governed by the first two terms of Eq.~(\ref{Lagrangian}), of the field 
$\varphi$, which here corresponds to triplon propagation. The vertex marked 
by the solid circle represents the bare interaction, the third term of 
Eq.~(\ref{Lagrangian}), whose coefficient, $\alpha$, is the perturbative 
parameter. The one-loop corrections to $\Gamma^{(4)}$ and $\chi_{\varphi\varphi}$ are equivalent to retaining next-to-leading-order terms in $\alpha$. 
For the expansion of $\Gamma^{(4)}$ this implies $\alpha^2$ terms, which are 
contained in the three distinct loop diagrams (the Mandelstam {\it s}, {\it t}, 
and {\it u} channels) in panel (a). For the expansion of $\chi_{\varphi\varphi}$ this is the order-$\alpha$ loop diagram in panel (b), to which we refer 
as the self-energy, $\Sigma$. The perturbative RG treatment of $\Gamma^{(4)}$ 
and $\chi_{\varphi\varphi}$ (Sec.~\ref{QFToneloop}) determines the running 
coupling constant (\ref{RunningAlpha}) and the running mass (\ref{RunningMass}) 
of the QFT description.}
\label{figvert}
\end{figure}
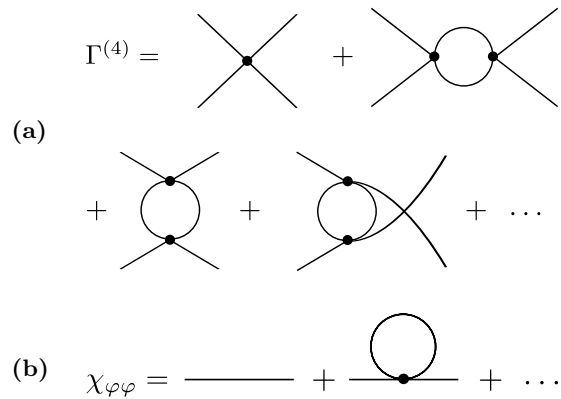

This mean-field analysis accounts for neither quantum nor thermal 
fluctuations. These we include in the present analysis at one-loop order, 
meaning that we consider contributions from the vertex and self-energy 
diagrams shown in Fig.~\ref{figvert}. To provide a self-contained treatment, 
in Sec.~\ref{QFToneloop} we demonstrate the procedure for the RG resummation, 
by which we obtain the one-loop quantum and thermal corrections that are 
central to the analysis of Secs.~\ref{Results} and \ref{Decay}.

\subsection{Quantum field theory: One-loop corrections} \label{QFToneloop}

The purpose of the present study is to obtain explicit expressions for the 
order parameter, excitation gaps, and N\'eel temperature, and hence all of 
their critical exponents, within the one-loop RG treatment of the QFT. To 
derive an analytic expression for the N\'eel temperature on the same footing 
as the zero-temperature quantities, it is necessary also to extend the 
analysis to finite temperatures. We take $J = 1$ as the unit of energy and 
set the fundamental constants $\hbar = 1$ and $k_{\rm B} = 1$. In the QFT, 
$\omega_k = \sqrt{c^2 k^2 + m^2}$ is the energy of a magnetic excitation at 
momentum (wave vector) ${\bm k}$, which is measured from the antiferromagnetic 
ordering wave vector, ${\bm Q} = (\pi,\pi,\pi)$. This matches the low-energy 
form of gapped or gapless spin excitations in the starting Hamiltonian 
(\ref{Hamiltonian}), while details of the higher-lying band excitations 
are not relevant to QFT. 

\subsubsection{Renormalization and Running Coupling}

We generalize the 3+1D QFT to an O($N$) theory and demonstrate the 
renormalization of the coupling constant, $\alpha$, of the Lagrangian 
(\ref{Lagrangian}) by considering the quantum disordered phase ($g > g_c$). 
The requirements of energetic scale-invariance give rise to the RG treatment 
of the QFT. We illustrate the RG process by evaluating the one-loop correction 
to the four-point vertex shown in Fig.~\ref{figvert}, 
\begin{align}
\notag \Gamma^{(4)} & = 6 \alpha + 6 (N+8)\, \alpha^2 \int^{\Lambda_c}_{\Lambda} 
\frac{d^4k}{(2\pi)^4 c^3} \frac{i}{(k^2 - m^2)^2} \\ \label{fourpoint}
& = 6 \alpha - \frac{6(N+8)\,\alpha^2}{8 \pi^2 c^3} \ln \left[ \frac{\Lambda_c} 
{\Lambda} \right]
\end{align}
if $\Lambda \ge m$. Here $k^2 = \omega^2 - c^2 {\bm k}^2$ is the square of the 
four-momentum, the factor of $1/c^3$ arises from rescaling the integration 
measure, and $m$ serves as the lower bound of the infrared cutoff, $\Lambda$. 
The first term in Eq.~(\ref{fourpoint}) corresponds to the first diagram in 
the perturbative series for $\Gamma^{(4)}$ represented in Fig.~\ref{figvert}(a) 
and the second to the three O($\alpha^2$) diagrams. A detailed discussion of 
the four-point vertex may be found in Ref.~\cite{Kleinert2001}; the common 
factor of 6 is absorbed in constants of proportionality and the universal 
factor of $(N+8)$ accounts for the number of inequivalent diagrams 
contributing at this order. 

The primary purpose of renormalization is to control the ultraviolet 
divergence, which is expressed in Eq.~(\ref{fourpoint}) by $\Lambda_c$; in 
a lattice problem such as the double-cubic model, the ultraviolet momentum 
cut-off is the inverse lattice spacing. The beta function of the RG flow is 
obtained from the Callan-Symanzik equation,
\begin{align}
\notag \left[ \frac{d}{d\ln(\Lambda_c/\Lambda)} + \beta(\alpha) 
\frac{d}{d\alpha} \right] \Gamma^{(4)} & = 0 
\end{align}
whence
\begin{align}
\notag \beta(\alpha) & = \frac{(N+8) \alpha^2}{8 \pi^2 c^3} \\ \notag
\frac{d\alpha}{d\ln(\Lambda_0/\Lambda)} & = - \frac{(N+8)\alpha^2}{8 \pi^2 
c^3} \\ \label{RunningAlpha} \alpha(\Lambda) \equiv \alpha_{\Lambda} & = 
\frac{\alpha_0}{1 + \frac{(N+8)}{8 \pi^2 c^3} \alpha_0 \ln(\Lambda_0/\Lambda)}.
\end{align}
This demonstrates explicitly how the RG procedure removes the dependence on 
$\Lambda_c$ by introducing a normalization point, $\Lambda_0$, which is a 
parameter that can be fixed by optimizing the fit to the starting model. 
The RG equations nevertheless retain a dependence on the infrared energy 
scale, $\Lambda$, which is the actual energy scale of the QFT and is set 
by the physical energy scale of the system. Because this is either the mass 
(gap) of the field $\varphi$ or the ordering temperature, both of which may 
vanish within the range of parameters covered by the QFT, $\Lambda$ is known 
as the ``running'' energy scale. In the renormalization process, this running 
is absorbed into the coupling constant, $\alpha \rightarrow \alpha_\Lambda$, 
giving it the dependence on $\Lambda$ specified in Eq.~(\ref{RunningAlpha}), 
i.e.~the running coupling constant, $\alpha_\Lambda$, is defined in terms of 
the constant $\alpha_0 \equiv \alpha(\Lambda_0)$.

The running of $\alpha_\Lambda$ as a logarithmic function of the infrared 
energy scale is an important and generic property of this type of QFT at the 
upper critical dimension, $d = 3 + 1$. As will become clear below, the static 
and dynamic observables derived from the QFT all depend explicitly on 
$\alpha_\Lambda$, and hence also depend logarithmically on the energy scale 
$\Lambda$. It is precisely this logarithmic dependence in the QFT that 
produces the scaling forms of Eqs.~(\ref{observablesms})-(\ref{observablesTN}), 
which were observed in the QMC simulations, and we will demonstrate this 
explicitly in Eqs.~(\ref{order})-(\ref{TNeelTriplon}). A further essential 
property of Eq.~(\ref{RunningAlpha}) that $\alpha_\Lambda \rightarrow 0$ as 
$\Lambda \rightarrow 0$, which is a statement that at the QCP, where all 
energy scales vanish (hence $\Lambda \rightarrow 0$), the running coupling 
vanishes. Thus one expects a weak-coupling theory in the vicinity of the QCP, 
a result important both for its inherent physical content and because it 
justifies the use of a one-loop perturbative treatment.

\subsubsection{Self-Energy in the Disordered Phase}

We now consider the renormalization of the triplon gap in the disordered phase. 
The first perturbative correction to the gap energy is given by the one-loop 
self-energy shown in Fig.~\ref{figvert}(b), which we separate into its 
zero-point and thermal contributions 
\begin{align}
\label{oneloop}
\notag \Sigma(\Delta,T) & = (N+2) \, \alpha_{\Lambda} \sum_{\bf k} \frac{1}
{\omega_{\bm k}} \left[ \frac{1}{2} + \frac{1}{e^{{\omega}_{\bm k}/T} - 1} 
\right] \\ & = (N+2) \, \alpha_{\Lambda} \int \frac{d^3k}{(2\pi)^3} \frac{1}
{2\omega_{\bm k}} \\ \notag & \;\;\;\;\;\;\;\; + (N+2) \, \alpha_{\Lambda} \int 
\frac{d^3k}{(2\pi)^3} \frac{1}{\omega_{\bm k}} \frac{1}{(e^{{\omega}_{\bm k}/T}
 - 1)}.
\end{align}
Because corrections to the response function are multiplicative with the 
four-point vertices, the relevant coupling constant is the running coupling, 
$\alpha_{\Lambda}$. The notation is chosen to clarify that the triplon gap and 
the self-energy are determined self-consistently,
\begin{align}
\label{masszeroT}
\Delta^2 (\delta g, T) = m^2(\delta g) + \Sigma(\Delta,T).
\end{align}
To analyze the renormalization of the bare mass, we consider the case of zero 
temperature, where only the first term of Eq.~(\ref{oneloop}) contributes. The 
leading contributions to the response function of Fig.~\ref{figvert}(b), which 
are responsible for the logarithmic corrections, are obtained by summing the 
Dyson series, and hence the inverse response function can be expressed in 
the closed form 
\begin{align}
\notag \chi^{-1}_{\varphi\varphi} (p) & = p^2 - m^2 - \Sigma(m, T = 0) \\
\notag & = p^2 - m^2 - (N+2) \alpha_{\Lambda} \int^{\Lambda_c}_{0} \frac{d^3k}
{(2\pi)^3} \frac{1}{2\sqrt{c^2 k^2 + m^2}} \\ & = p^2 - m^2 + \frac{(N+2)
\alpha_{\Lambda}}{8 \pi^2 c^3} m^2 \ln \left( \frac{\Lambda_c}{m} \right) \! ,
\label{masslog}
\end{align}
where $p$ is the external four-momentum and $p^2 = \omega^2 - {\bm p}^2$. We 
apply the Callan-Symanzik procedure to obtain the beta function for the mass, 
for which we again substitute $\Lambda$ in place of $m$ as the lower energy 
cut-off in the logarithm (\ref{masslog}). From
\begin{align}
\notag 0 & = \left[ \frac{d}{d\ln(\Lambda_c/\Lambda)} + \beta_m (\Lambda) 
\frac{d}{dm^2} \right] \chi_{\varphi\varphi} (p = 0) \\ \notag
\beta_m (\Lambda) & = \frac{(N+2) \alpha_{\Lambda} m^2}{8 \pi^2 c^3} \\ \notag 
\frac{d m^2}{d \ln(\Lambda_0/\Lambda)} & = - \frac{(N+2) \alpha_{\Lambda} 
m^2}{8 \pi^2 c^3} \\ \notag \frac{d \ln (m^2)}{d \ln(\Lambda_0/\Lambda)} & =
 - \left( \frac{N+2}{N+8} \right) \frac{\frac{N+8}{8 \pi^2 c^3} \alpha_0}{1
 + \frac{(N+8)}{8 \pi^2 c^3} \alpha_0 \ln (\Lambda_0/\Lambda)} 
\end{align}
we obtain 
\begin{align}\label{RunningMass} 
m^2_{\Lambda} & = m_0^2 \left( \frac{\alpha_\Lambda}{\alpha_0} \right)^{\frac{N+2}{N+8}},
\end{align}
and thus the triplon gap at zero temperature is given by $\Delta_t \equiv 
m_\Lambda$, which specifies its critical exponent [Eq.~(\ref{observablest})] as 
\begin{equation}
\beta_2 = \frac{N+2}{2(N+8)}. 
\end{equation} 
We defer the explicit rearrangement of Eq.~(\ref{RunningMass}) in the form of 
Eq.~(\ref{observablest}) to Sec.~\ref{Model}D.

The corrections at finite temperatures may be computed from the second term 
of Eq.~(\ref{oneloop}), the thermal contribution to the one-loop self-energy.
Without presenting an explicit evaluation, we state that this does not change 
the form of the running coupling (\ref{RunningAlpha}) and hence does not change 
the form of the running mass (\ref{RunningMass}), but it does present a 
possible change to the infrared cutoff, from $\Lambda = \Delta_t(\delta g)$ 
to $\Lambda = {\rm Max} \{\Delta_t(\delta g,T),T\}$. We collect the 
scale-dependence contained in Eq.~(\ref{RunningMass}) into a gap expression 
of the form 
\begin{align}
\label{DeltaT}
\Delta^2_t(\delta g, T, \Lambda) & = \gamma^2 \delta g \left[\frac{\alpha
_{\Lambda}}{\alpha_0} \right]^{\frac{N+2}{N+8}} \\ \notag &  \;\;\;\;\;\; + (N+2) \, 
\alpha_{\Lambda} \sum_{\bf k} \frac{1}{\omega_{\bm k}} \frac{1}{e^{{\omega}_{\bm k}/T}
 - 1}.
\end{align}

\subsubsection{Self-Energy in the Ordered Phase}

We conclude our overview of one-loop corrections by considering renormalization 
in the ordered phase, which is induced by the spontaneous breaking of the 
O($N$) symmetry when $g < g_c$. Calculating perturbative corrections to the 
Higgs gap, and hence obtaining the correct critical exponents, is a delicate 
task in the ordered phase because the results must preserve the Goldstone 
theorem at each order in $\alpha$. The Goldstone theorem is a direct result 
of the remaining O($N-1$) symmetry and dictates that the Goldstone modes 
must remain massless even after perturbative corrections. To outline the 
appropriate procedure for computing corrections to the order parameter and 
the Higgs gap, we consider the general case of finite temperature, which is 
required to obtain $T_N$. 

We write the field in the Lagrangian (\ref{Lagrangian}) as $\vec{\varphi}
 = (\varphi_c + \sigma, \vec{\pi})$, where the minimum of the potential 
(expectation value of the finite static field) is $\varphi_c$ and the field 
oscillations about this shifted minimum are the $N-1$ Goldstone modes, 
$\vec{\pi}$, and the gapped Higgs mode, $\sigma$. The effective potential, 
${\cal V}$, due to the non-derivative terms in Eq.~(\ref{Lagrangian}), 
when expanded about $\varphi_c$, are 
\begin{align}\label{elop}
{\cal V} & = - {\textstyle \frac{1}{2}} |m^2| (\varphi_c + \sigma, \vec{\pi})^2
 + {\textstyle \frac{1}{4}} \alpha \left[ (\varphi_c + \sigma, \vec{\pi})^2 
\right]^2.
\end{align}
The two conditions 
\begin{align}
\label{conditions}
\frac{d{\cal V}}{d\vec{\varphi}} \Big|_{\varphi_c} = 0 \ \ \ \ \ \ \text{and} 
\ \ \ \ \ \ \frac{d^2{\cal V}}{d\vec{\pi}^2} \Big|_{\varphi_c} = 0
\end{align}
must hold simultaneously to ensure that $\varphi_c$ is indeed the minimum of 
the potential and that, to any order in $\alpha$, the perturbations respect 
the O($N-1$) symmetry and so preserve the Goldstone theorem. Because we have 
already obtained the universal scale dependence of $\alpha_{\Lambda}$, and 
hence of $m_{\Lambda}$, there is no need to repeat the Callan-Symanzik RG 
procedure, but it remains to treat the thermal contributions more explicitly. 
By satisfying Eq.~(\ref{conditions}) at one-loop order we obtain
\begin{align}
\label{tadpole}
\notag \frac{d{\cal V}}{d\vec{\varphi}} \Big|_{\varphi_c} & = \alpha_{\Lambda} 
\varphi_c^2 - |m_{\Lambda}^2| +(N-1) \alpha_{\Lambda} \sum_{\bf k} \frac{1/(ck)}
{e^{ck/T} - 1} \\ & \;\;\;\;\;\;\;\; + 3 \alpha_{\Lambda} \sum_{\bf k} 
\frac{1/{\omega}_{\bm k}}{e^{{\omega}_{\bm k}/T} - 1} \;\; = \;\; 0, 
\end{align}
whence
\begin{align}
\label{phi_corrections}
\varphi_c^2 & = \frac{|m_{\Lambda}^2|}{\alpha_{\Lambda}} - (N \! - \! 1) 
\! \sum_{\bf k} \! \frac{1/(ck)}{e^{ck/T} \! - \! 1} - 3 \! \sum_{\bf k} \! 
\frac{1/{\omega}_{\bm k}}{e^{\omega_{\bm k}/T} \! - \! 1}.
\end{align}
Here we have separated the thermal contributions to the self-energy into two 
summations, the first with a (massless) Goldstone propagator in the loop and 
the second with a Higgs propagator whose mass is contained in 
$\omega^2_{\bm k} = c^2 k^2 + \Delta_H(\delta g,T)^2$. This separation 
is discussed in greater detail in Sec.~\ref{Decay}, where it is represented 
explicitly in Fig.~\ref{Vector}. The Higgs gap is given at one-loop order by 
\begin{align}
\notag \Delta_H^2 & = 3 \alpha_{\Lambda} \varphi_c^2 - |m_{\Lambda}^2| + (N-1) 
\alpha_{\Lambda} \sum_{\bf k} \frac{1/(ck)}{e^{ck/T} - 1} \\ \notag & \;\;\;\;
\;\;\;\; + 3 \alpha_{\Lambda} \sum_{\bf k} \frac{1/{\omega}_{\bm k}}{e^{\omega_{\bm k}/T}
 - 1} \\
\notag & = 2 |m_{\Lambda}|^2 - 2 (N-1) \alpha_{\Lambda} \sum_{\bf k} \frac{1/(ck)} 
{e^{ck/T} - 1} \\ \label{HiggsT} & \;\;\;\;\;\;\;\; - 6 \alpha_{\Lambda} 
\sum_{\bf k} \frac{1/{\omega}_{\bm k}}{e^{{\omega}_{\bm k}/T} - 1} \\ \label{Higgsexp}
& = 2 \alpha_{\Lambda} \varphi_c^2 + {\rm O}(\alpha^2),
\end{align}
where we have made use of Eq.~(\ref{phi_corrections}) at both steps. It is 
evident from Eq.~(\ref{phi_corrections}), where the latter two terms have 
no explicit dependence on a running quantity, that the critical exponent of 
the order parameter is $\beta_1 = \beta_2 - 1/2 = 3/(N+8)$ and from 
Eq.~(\ref{Higgsexp}) that for the Higgs gap it is $\beta_3 = \beta_2$.

Finally, the N\'eel temperature can be calculated by approaching the QCP 
from the ordered phase and solving Eq.~(\ref{HiggsT}) with $\Delta_H (\delta 
g,T_N) = 0$ to obtain 
\begin{align}
\label{Tneel}
T_N^2 (\delta g) & = \frac{12 \gamma^2 |\delta g| c^3}{(N+2) \alpha_0} 
\left[ \frac{\alpha_0}{\alpha_{\Lambda}} \right]^{\frac{6}{N+8}}.
\end{align}
Approaching from the disordered phase and solving Eq.~(\ref{DeltaT}) with 
$\Delta_t (\delta g, T_N) = 0$ gives an identical result. It is clear that 
the critical exponent $\beta_4 = \beta_1$. 

\subsection{QFT observables}

For comparison with the QMC observables in 
Eqs.~(\ref{observablesms})-(\ref{observablesTN}), we gather the four 
quantities derived from the one-loop RG calculations of Sec.~\ref{QFToneloop} 
in the form \begin{widetext}
\begin{align}
\label{order}
\varphi_c^2 (\delta g) & = \frac{\gamma^2|\delta g|}{\alpha_0} \left[ 
\frac{\alpha_0}{\alpha_{\Delta}} \right]^{\frac{6}{N+8}} & & \hspace{-1.2cm}
 = \frac{\gamma^2}{\alpha_0 g_c} \left( \frac{16 \pi^2 c^3}{(N+8)\alpha_0} 
\right)^{\frac{-6}{N+8}} |g - g_c| \left| \ln \left( \frac{|g - g_c|}{\tilde{b}_1} 
\right) \right|^{\frac{6}{N+8}}, \\
\label{gap}
\Delta_t^2 (\delta g) & = \gamma^2 |\delta g| \left[ \frac{\alpha_{\Delta}}
{\alpha_0} \right]^{\frac{N+2}{N+8}} & & \hspace{-1.2cm} = \frac{\gamma^2}{g_c}
\left( \frac{16 \pi^2 c^3}{(N+8) \alpha_0} \right)^\frac{N+2}{N+8} |g - g_c| 
\left| \ln \left( \frac{|g - g_c|}{\tilde{b}_2} \right) \right|^{-\frac{N+2}{N+8}}, 
\\ \label{HiggsGap}
\Delta_H^2 (\delta g) & = 2 \gamma^2 |\delta g| \left[ \frac{\alpha_{\Delta}}
{\alpha_0} \right]^{\frac{N+2}{N+8}} & & \hspace{-1.2cm} = 2 \frac{\gamma^2}{g_c}
\left( \frac{16 \pi^2 c^3}{(N+8) \alpha_0} \right)^\frac{N+2}{N+8} |g - g_c| 
\left| \ln \left( \frac{|g - g_c|}{\tilde{b}_3} \right) \right|^{-\frac{N+2}{N+8}}, 
\\ \label{TNeelTriplon}
T_N(\delta g)^2 & = \frac{12\gamma^2 |\delta g| c^3}{(N+2)\alpha_0} \left[ 
\frac{\alpha_0}{\alpha_{T_N}} \right]^{\frac{6}{N+8}} & & \hspace{-1.2cm} = 
\frac{12 \gamma^2 c^3}{(N+2) \alpha_0 g_c} \left( \frac{16 \pi^2 c^3}{(N+8) 
\alpha_0} \right)^{\frac{-6}{N+8}} |g - g_c| \left| \ln \left( \frac{|g - g_c|}
{\tilde{b}_4} \right) \right|^{\frac{6}{N+8}}.
\end{align}
\end{widetext}
Here $g_c$ and $c$ are constants of the double-cubic system and $N = 3$.
The logarithmic dependence of the right-hand side on $|\delta g|$ enters due 
to the logarithmic scale dependence of the running coupling constant given 
in Eq.~(\ref{RunningAlpha}), from which the quantities $\alpha_\Delta$ and 
$\alpha_{T_N}$ are obtained by setting $\Lambda = \max\{\Delta_{t},\Delta_{H}/
\sqrt{2},T\}$ to the largest energy scale in the system. Here we take the 
running scale to be $\Delta_{H}/\sqrt{2} = |\Delta_t|$ for the three quantities 
$m_s(\delta g)$, $\Delta_t(\delta g)$, and $\Delta_H(\delta g)$. 

\begin{figure*}[t]
\includegraphics[width=0.435\textwidth,clip]{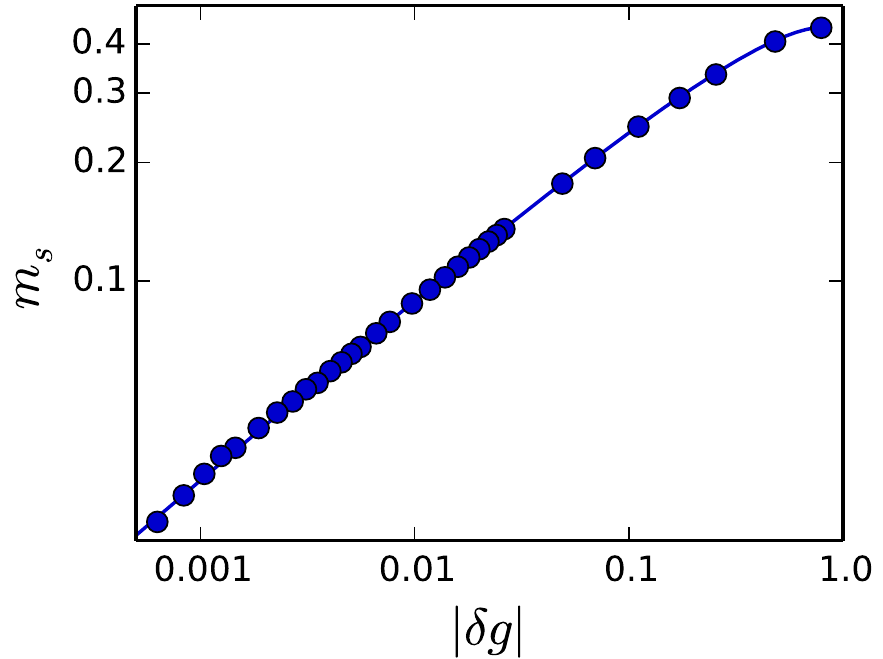}
\hspace{1.cm}\vspace{0.25cm}
\includegraphics[width=0.435\textwidth,clip]{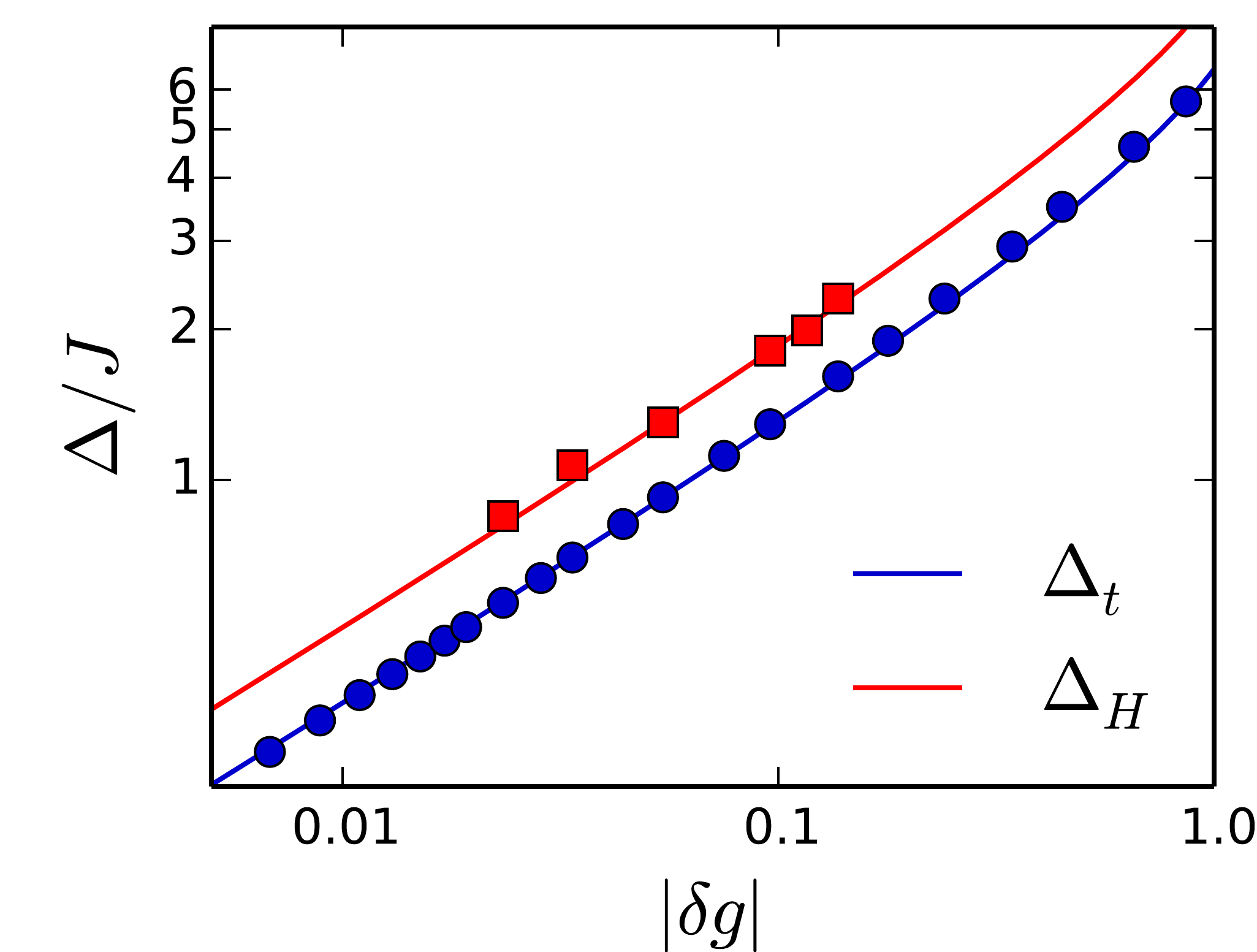} 
\includegraphics[width=0.445\textwidth,clip]{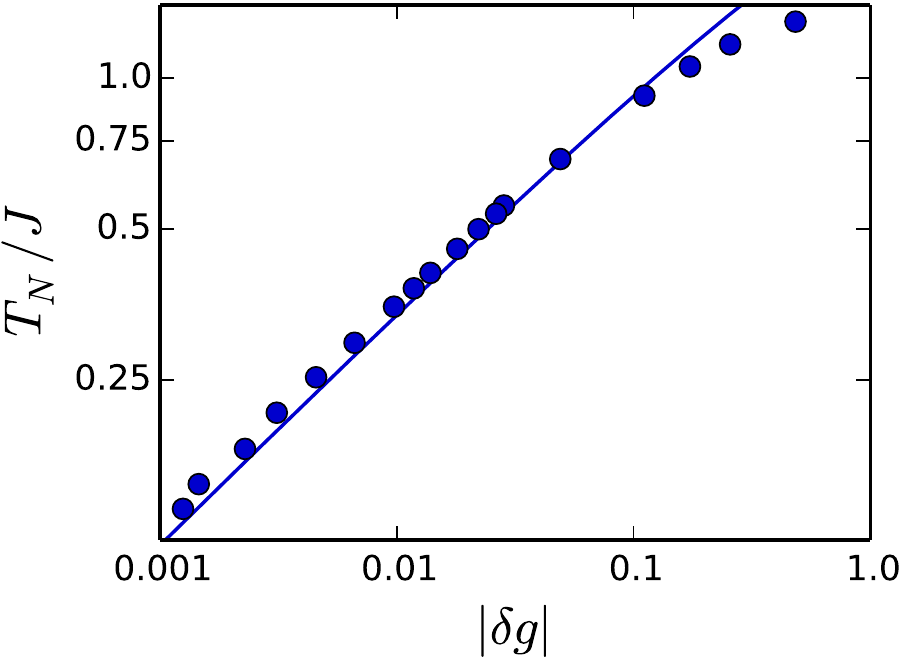} 
\hspace{0.8cm}
\includegraphics[width=0.445\textwidth,clip]{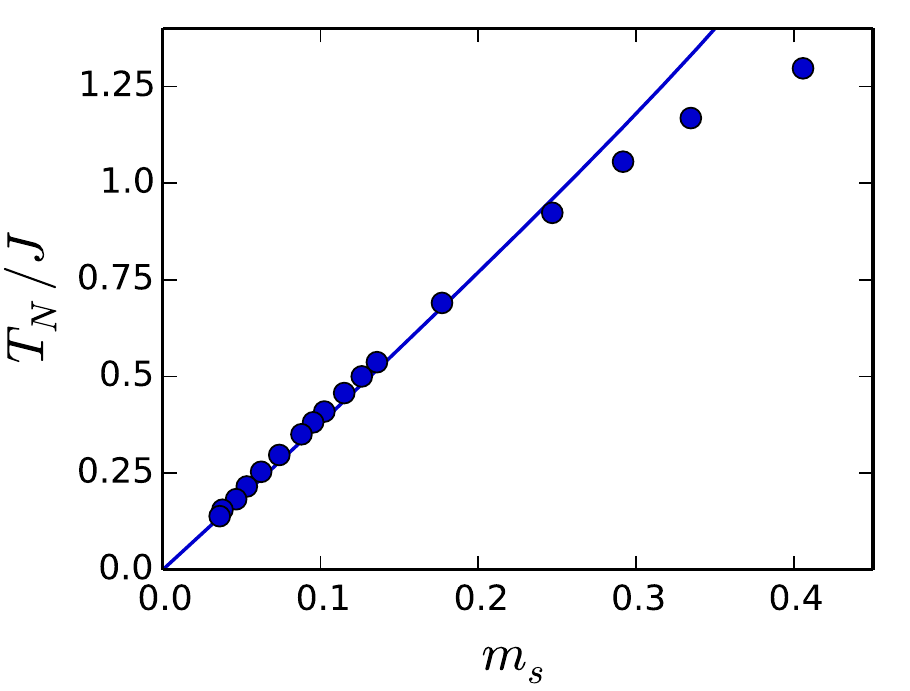}
\begin{picture}(0,0) 
\put(-177,330){\text{{\bf (b)}}} 
\put(-177,150){\text{{\bf (d)}}} 
\put(-435,330){\text{{\bf (a)}}} 
\put(-435,150){\text{{\bf (c)}}} 
\end{picture}
\caption{Static and dynamic observables shown as a function of the separation, 
$|\delta g|$, from the QCP. Discrete points are extrapolated QMC data and 
lines are drawn from QFT fitting. (a) Staggered magnetization, $m_s$, in the 
ordered phase ($g < g_c$); QFT fit from Eq.~(\ref{order}) with $m_s = 
\Upsilon^{-1} \varphi_c$. (b) Triplon gap, $\Delta_t$, in the disordered phase 
($g > g_c$) and Higgs gap, $\Delta_H$, in the ordered phase; QFT fits from 
Eqs.~(\ref{gap}) and (\ref{HiggsGap}). (c) N\'eel temperature, $T_N$, in the 
ordered phase; QFT fit from Eq.~(\ref{TNeelTriplon}). (d) $T_N$ compared to 
$m_s$, with $\delta g$ as the implicit parameter.}
\label{fitsgap}
\end{figure*}

Explicitly, the dependence of $\alpha_\Lambda$ on the separation, $\delta g$, 
from the QCP is given by 
\begin{align}
\label{alpha_alt}
\alpha_\Lambda(\delta g) & = \frac{16 \pi^2 c^3}{(N+8)} \left| \, \ln \left( 
\frac{|g - g_c|}{\tilde{b}_i} \right) \right|^{-\frac{N+2}{N+8}},
\end{align}
where 
\begin{align}
\label{b}
\notag \tilde{b}_1 & = \tilde{b}_2 = \tilde{b}_3 = \frac{g_c\Lambda_0^2}
{\gamma^2} e^{\frac{16\pi^2}{(N+8)\alpha_0}}, \\
\tilde{b}_4 & = \frac{(N+2)\alpha_0 g_c \Lambda_0^2}{12 c^3 \gamma^2}
e^{\frac{16\pi^2}{(N+8)\alpha_0}}.
\end{align}
Thus the three zero-temperature coefficients $\tilde{b}_{1,2,3}$ are equal, 
but different from $\tilde{b}_4$ determined on the N\'eel-temperature 
curve. We note that an exact derivation of coefficients appearing within 
the logarithms is beyond the scope of one-loop RG and would require higher 
loop corrections. 

It is important to stress that the running coupling is a function of the 
energy-scale ratio $\Lambda_0/\Lambda$ that is determined uniquely by 
Eq.~(\ref{RunningAlpha}). However, when parameterized in terms of $\delta g$ 
[Eq.~(\ref{alpha_alt})] it is necessary to include the constants $\tilde{b}_i$ 
to account for the different possible dependences of $\Lambda_0/\Lambda$ on 
$|g - g_c|$. Equation (\ref{alpha_alt}) serves three purposes in the present 
context. First, it allows for a simple conversion between the running coupling 
constant of QFT and the logarithmic scaling forms used widely in condensed 
matter \cite{Zinn-Justin}. Second, it demonstrates how QFT specifies the 
closely related functional forms of all four observables. Third, it shows 
explicitly how the five fundamental parameters of the QFT give a unique and 
quantitative determination of these observables; alternatively stated, the 
parameters $\{a_i, b_i\}$ and exponents $\{\nu_i,\beta_i\}$ required to fit 
the numerical data using Eqs.~(\ref{observablesms})-(\ref{observablesTN}) 
are obtained directly.

\section{Static and dynamic observables}\label{Results}

\subsection{Fitting parameters}

Here we present the results obtained by fitting the QMC data for the 
staggered magnetization, the triplon and Higgs excitation gaps, and the 
N\'eel temperature \cite{Qin2015,Qin2017} using the QFT expressions of 
Eqs.~(\ref{order})-(\ref{TNeelTriplon}) and extract the numerical values 
of the remaining free parameters. The constants $g_c = 4.83704$ and $c = 
2.365$ for the dimerized QAF on the double-cubic lattice are taken directly 
from QMC. Because the QFT framework presents a means of connecting sets of 
observables that are determined independently by QMC, we perform a complete 
fit to all data sets simultaneously. However, to do this in a reliable 
manner, the influence of different QMC points and of different datasets
should be weighted according to their statistical reliability. Following 
the discussion in Sec.~\ref{Model}A, we weight the QMC datasets in the 
order $m_s > \Delta_t > \Delta_H > T_N$. As explained in more detail in 
Sec.~\ref{Results}B, we give equal weight to all QMC data points in each 
set with $|\delta g| \le 0.2$ and none to those at higher $|\delta g|$. 
The results are shown in Fig.~\ref{fitsgap}.

These fits contain two adjustable parameters, which can be expressed as 
the mass proportionality factor $\gamma$ and the ratio $\alpha_0/(8\pi c^3)$. 
It is important to note that the choice of the normalization point, 
$\Lambda_0$, is arbitrary, and affects directly the value of $\alpha_0$;
because $\alpha_0 \equiv \alpha_{\Lambda_0}$, any other choice of the 
normalization point, $\Lambda'_0 \neq \Lambda_0$, simply redefines 
$\alpha_{\Lambda'_0} \equiv \alpha'_0$. Here we make the explicit choice 
$\Lambda_0 = 0.915 J$, based on the criterion $\Delta_t(\Lambda_0) = 
\Lambda_0$, which proves to be convenient for the comparison with a 
bond-operator description (Sec.~\ref{derivation}). With this choice, the 
adjustable parameters are found to be
\begin{align}
\label{bestfit1}
\alpha_0/(8\pi c^3) = 0.175, \ \ \gamma = 3.95 J,
\end{align}
and hence
\begin{align}
\label{bestfit2}
{\tilde b}_{1,2,3} = 6.78, \ \ {\tilde b}_4 = 12.43.
\end{align}
Finally, an explicit relationship between the QFT order parameter, $\varphi_c$, 
and $m_s$ determined directly from QMC lies beyond the reach of QFT. We assume 
the relation
\begin{align}
\label{upsilon}
\varphi_c & = \Upsilon m_s,
 \end{align}
and obtain $\Upsilon = 0.65$ for the constant of proportionality. In 
Sec.~\ref{derivation} we justify the assumption of linearity and provide an 
analytic expression for $\Upsilon$ based on the bond-operator technique. 

Figures \ref{fitsgap}(a), \ref{fitsgap}(b), and \ref{fitsgap}(c) show 
respectively our fits to $m_s$ (\ref{order}), $\Delta_t$ (\ref{gap}) and 
$\Delta_H$ (\ref{HiggsGap}), and $T_N$ (\ref{TNeelTriplon}), which were made 
using the parameters of Eqs.~(\ref{bestfit1}) and (\ref{bestfit2}). The 
logarithmic axes are chosen to highlight the multiplicative corrections as 
departures from the straight-line form of the mean-field exponents. Our major 
conclusion is the remarkable agreement between QMC and QFT, which demonstrates 
clearly that QFT, with a single set of parameters, is capable of providing a 
quantitative description, and hence a unification, of static and dynamic 
observables. This procedure also demonstrates once again, to high precision, 
the validity of the theoretical predictions of the O(3) QFT. 

We comment that our fits in Fig.~\ref{fitsgap} are not identical to those 
of Ref.~\cite{Qin2015}. In the QMC study, the fits were found to be very 
insensitive to the values of the parameters ${\tilde b}_i$, which were set to 
$g_c$. In the QFT analysis, we gain both deeper insight into these parameters 
and a means of fixing them through constants to which the fits are more 
sensitive [Eq.~({\ref{b})]. The ${\tilde b}_i$ values we obtain account for 
the minor quantitive differences between the fits, although we also did not 
implement an error-bar weighting as in Ref.~\cite{Qin2015}. The QFT analysis 
also affords extra insight into the linearity of $T_N$ and $m_s$, which is 
shown in Fig.~\ref{fitsgap}(d). First observed numerically in 
Ref.~\cite{Jin2012}, the almost exact linearity of the two parameters was 
studied in detail in Ref.~\cite{Qin2015}, where it was found that the two 
have the same logarithmic corrections; a scaling argument was formulated 
in support of this result, which has recently been observed again in a 
similar context \cite{Tan2017}. From QFT it is clear immediately that $m_s$ 
(\ref{order}) and $T_N$ (\ref{TNeelTriplon}) have multiplicative logarithmic 
corrections with the same exponent, illustrating again the unifying nature 
of the analysis. However, the arguments of the logarithms are not identical, 
due to the different cut-off energy scales, which are reflected in the 
different constants ${\tilde b}_1$ and ${\tilde b}_4$, and this is why 
the QFT fit in Fig.~\ref{fitsgap}(d) is not in fact a completely straight 
line at large $|\delta g|$. 

\subsection{Quantum critical regime}

A key question in the theory of quantum critical systems is to understand 
the width of the quantum critical regime [Fig.~\ref{phase}(b)], by which is 
meant the region of the phase diagram where the predicted quantum critical 
scaling forms [Eqs.~(\ref{order})-(\ref{TNeelTriplon})] remain applicable. 
The standard arguments of perturbative one-loop RG contain no such 
information, and cannot guarantee that the quantum critical regime is more 
than an asymptotic concept reached only when $|\delta g| \rightarrow 0$. 
Thus the width of this regime was referred to in Ref.~\cite{Qin2015} as 
one of the nonuniversal constants of the system and it may be regarded as 
something of a surprise that quantum critical scaling was found in the QMC 
data over the rather broad range $|\delta g| \leq 0.2$. This estimate was 
obtained using the scaling forms of Eqs.~(\ref{observablesms}) and 
(\ref{observablesTN}), which make no explicit reference to the running 
coupling constant, $\alpha_\Lambda$. Hence one may ask whether this aspect 
of the QFT description provides additional insight into the width of the 
quantum critical regime. 

Within the one-loop RG treatment, the QFT results remain accurate while the 
running coupling remains small, i.e.~$\alpha_\Lambda/(8\pi c^3) \ll 1$. This 
criterion is independent of the numerical analysis leading to $|\delta g| 
\leq 0.2$ and applies to all four of the observables we consider, which 
again demonstrates the unifying aspects of the QFT description. An explicit 
evaluation of Eq.~(\ref{alpha_alt}) shows that $\alpha_\Lambda/(8\pi c^3) = 1$, 
the absolute upper bound on the applicability of one-loop RG as applied here, 
corresponds to $|\delta g| \approx 0.8$. Although one may debate the meaning 
of ``small'' relative to unity, it appears that the QMC estimate $|\delta g| 
\leq 0.2$ lies comfortably within the regime of validity of the QFT results. 

One may, however, ask whether it is possible that quantum critical scaling 
could be obeyed for $|\delta g| \lesssim 1$. The agreement between the QFT 
form and the QMC data for both the staggered magnetization and the triplon 
gap [Figs.~\ref{fitsgap}(a) and \ref{fitsgap}(b)], suggests that this may 
be the case. Here we comment again that such a level of agreement was not 
obtained in the initial analysis of the QMC data \cite{Qin2015}, where the 
constant $b_1$ was imposed rather than deduced. Although the QFT fits shown 
in Fig.~\ref{fitsgap} were performed by using only the QMC data in the 
range $|\delta g| \leq 0.2$ (Sec.~\ref{Results}A), this level of agreement 
demonstrates that the process we apply does not dictate the answer we 
obtain. This said, here we believe that the excellent agreement at the upper 
limit of the data range, $|\delta g| = 0.8$, is probably accidental. There 
are no theoretical grounds on which to expect quantum critical scaling over 
such a broad parameter regime. The QFT analysis states that the description
is not reliable by the time $\alpha_\Lambda/(8\pi c^3) = 1$. Further, the 
rather abrupt disagreement between QFT and QMC for $T_N(|\delta g|)$, which
sets in beyond $|\delta g| \approx 0.1$ [Fig.~\ref{fitsgap}(c)], suggests 
that the agreement is not global; this degree of mismatch cannot be ascribed 
to the lower accuracy of the QMC $T_N$ data compared to that of the $m_s$ 
data (Sec.~\ref{Model}A). Thus QFT tends to reinforce the QMC estimate that 
the width of the quantum critical regime is around $|\delta g| \leq 0.2$. 
Nevertheless, to the extent that the region beyond this limit is a crossover 
regime, detailed QMC and QFT studies of the double-cubic lattice would be an 
excellent means of probing crossover physics. 

\section{Results: Higgs decay width}\label{Decay}

The stability of the amplitude mode is a topic of crucial importance from 
the Standard Model to condensed matter and ultracold atoms. The broken 
symmetry of the ordered state, which establishes the massive Higgs mode, 
also ensures that Goldstone modes are ubiquitous, and with them a Higgs 
decay channel. Here we restrict our considerations to the line width arising 
due to Higgs decay processes in the 3D dimerized QAF. In the neutron 
scattering experiments on TlCuCl$_3$ \cite{Ruegg2008,Merchant2014}, the 
amplitude mode was found, in contrast to the triplon modes, to have an 
intrinsic line width, which varied with temperature and proximity to the 
QCP. 

Theoretically, the line width is extracted from a response function. For 
a system represented by a vector field, one may consider the response to 
vector or a scalar probe. In this sense, neutron scattering is a vector 
probe and the vector response function it provides is the dynamical 
spin-spin correlation function. The very recent dynamical QMC studies 
\cite{Qin2017,Lohofer2017} applied advanced SAC methods to the imaginary-time 
Green functions obtained from SSE QMC to provide numerical data for both 
the vector and scalar response functions of the double-cubic QAF. Perhaps 
self-evidently, this analysis is restricted to the ordered phase, where 
spontaneous decay of the gapped triplet mode is possible; in the disordered 
phase, the spontaneous decay of triplons is forbidden by a lack of available 
phase space \cite{ScammellWidths2017}. 

To discuss the decay of the Higgs mode at the upper critical dimension 
by QFT, we continue the analysis of the ordered phase begun in Sec.~IIC3. 
When the vector field is reexpressed with an explicit separation of the 
amplitude component, i.e.~$\vec{\varphi} = (\varphi_c + \sigma,\vec{\pi})$, 
the $\alpha \vec{\varphi}^4$ interaction term in Eq.~(\ref{elop}) takes the 
form
\begin{align}
\label{Lint}
\cal{V}_{\text{Int}} & = {\textstyle \frac{1}{4}} \alpha (\sigma^4 + \vec\pi^4
 + 2 \sigma^2 \vec\pi^2 + 4 \varphi_c \sigma^3 + 4 \varphi_c \sigma \vec{\pi}^2).
\end{align}
The final term, $\alpha \varphi_c \sigma \vec{\pi}^2$ contains the 
leading-order coupling of the Higgs and Goldstone modes, which enables 
the decay of the former. We analyze this process by calculating the vector 
and scalar response functions within the one-loop QFT framework of 
Sec.~\ref{QFToneloop}, using the parameters derived in Sec.~\ref{Results}. 

\begin{figure}[t]
\hspace{-0.3cm}{\includegraphics[width=0.4\textwidth,clip]{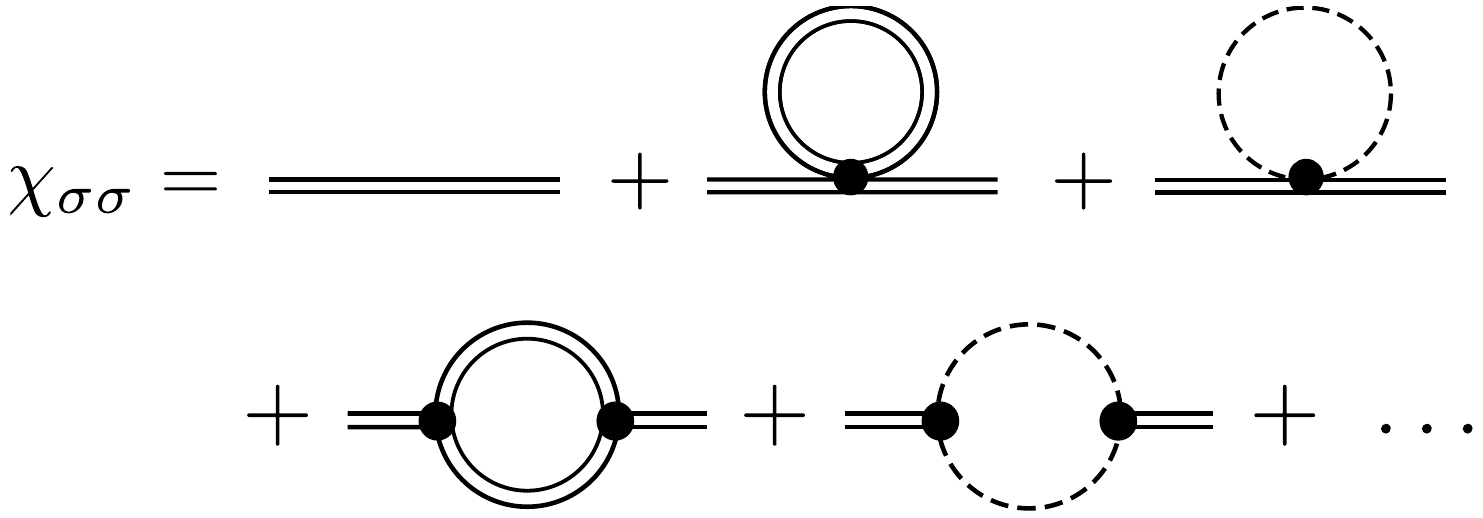}}
\caption{Diagrammatic expansion for the response function 
$\chi_{\sigma\sigma}$. The expansion is terminated at one-loop order, 
which corresponds to first order in $\alpha$. The double and dashed lines 
represent respectively the free propagation of the Higgs ($\sigma$) and 
Goldstone ($\vec{\pi}$) fields, obtained by setting $\vec\varphi
 = (\varphi_c + \sigma, \vec\pi)$ in Eq.~(\ref{Lagrangian}) and retaining 
terms to quadratic order in $\sigma$ and $\vec{\pi}$. Diagrams in the top 
line correspond to the quartic vertex terms, $\alpha \sigma^4$ and $\alpha 
\sigma^2 \vec{\pi}^2$ in (\ref{Lint}), and are clearly first-order in $\alpha$.
Diagrams in the bottom line correspond to the cubic vertex terms $\alpha 
\varphi_c \sigma^3$ and $\alpha \varphi_c \sigma \vec{\pi}^2$. Although each 
diagram is a product of two $\alpha$ vertices, the fact that the coefficient 
$\alpha^2 \varphi_c^2 = {\textstyle \frac{1}{2}} \alpha \Delta_H^2$, as shown 
in Eq.~(\ref{Higgsexp}) of Sec.~\ref{QFToneloop}, means that these terms 
remain first-order in $\alpha$. The evaluation of these diagrams is given 
by Eq.~(\ref{spectralvector}).}
\label{Vector}
\end{figure}

\subsection{Vector response function}

The vector response function is defined as $\chi_{\varphi\varphi} (p) = \langle 
\vec\varphi(p) \, \vec\varphi(0) \rangle$. In terms of the Higgs and Goldstone 
components,
\begin{align}
\label{chiexpansion}
\notag \chi_{\varphi\varphi} (p) & = \langle \sigma(p) \sigma(0) \rangle + (N - 1) 
\langle \pi(p) \pi(0) \rangle \\ & = \chi_{\sigma\sigma} (p) + (N - 1) \chi_{\pi\pi}
(p).
\end{align}
In this form, the vector response is summed over all components and corresponds 
to an unpolarized probe. In this sense Eq.~(\ref{chiexpansion}) is equivalent 
to the quantity calculated in the QMC simulations, which are performed on 
finite-size lattices with unbroken spin-rotation symmetry. We note that there 
are no cross components of the Higgs field and the order parameter, 
i.e.~$\chi_{\sigma\varphi_c} = 0$.

\begin{figure}[t]
\includegraphics[width=0.425\textwidth,clip]{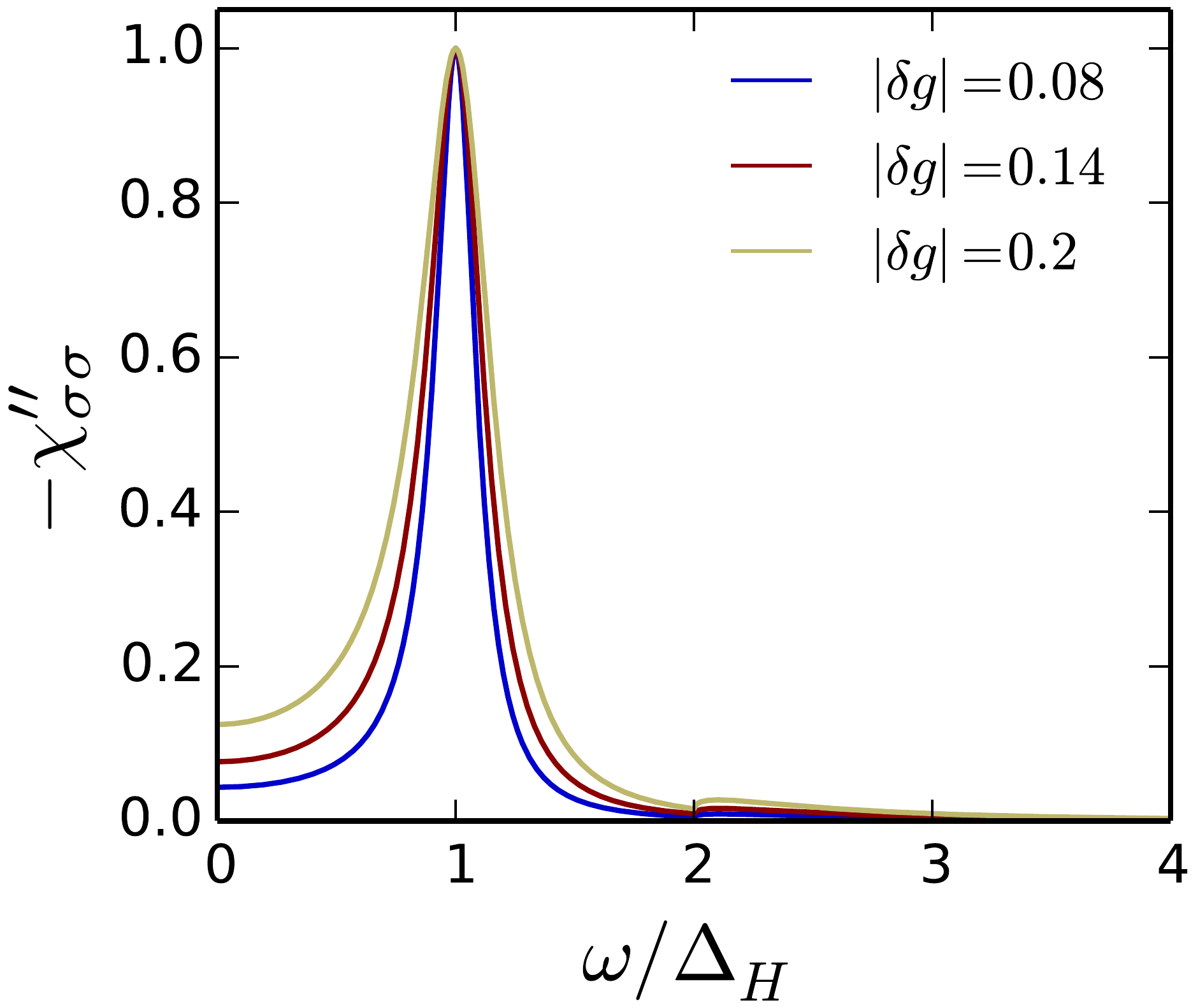}
\caption{Imaginary part of the vector response function, given by $-\chi''_{\sigma\sigma} (\omega)$, shown as a function of $\omega/\Delta_H$ at $\bm 
p = \bm 0$ and normalized to its maximum value. The curves correspond to 
different values, $|\delta g|$, of the coupling ratio relative to the QCP.}
\label{chiV}
\end{figure}

We compute the response function at first order in $\alpha$. The Goldstone 
contribution, $\chi_{\pi\pi} (p)$, has no one-loop corrections, which is a 
direct consequence of the Goldstone theorem demonstrated explicitly in 
Sec.~\ref{QFToneloop}, and hence 
\begin{align}
\chi_{\pi\pi}(p) & = \frac{1}{p^2 + i0}
\end{align}
where $i0$ in the denominator denotes the limiting imaginary part. The 
one-loop corrections to the Higgs component, represented in Fig.~\ref{Vector}, 
are finite, and their real part was treated explicitly in Eq.~(\ref{HiggsT}). 
In all of the equations to follow, the Higgs gap, $\Delta_H$, represents the 
one-loop renormalized value given in Eq.~(\ref{HiggsGap}) and it remains to 
evaluate the imaginary part of the one-loop corrections to the Higgs part of 
the response function. The first two loop diagrams on the right-hand side of 
Fig.~\ref{Vector} have purely real contributions, which are thus contained 
in the renormalized $\Delta_H$, and only the two terms on the lower line 
give imaginary contributions. These we label $\Pi_H(p)$ and $\Pi_G(p)$ to 
denote polarization loops with, respectively, with two Higgs and two 
Goldstone internal lines. Again their real parts have already been included 
in Eq.~(\ref{HiggsT}) and their imaginary parts, $\Pi''_H(p)$ and $\Pi''_G(p)$, 
give the result 
\begin{align}
\label{spectralvector}
\chi_{\sigma\sigma}(p) & = \frac{1}{p^2 - \Delta_H^2 - {\textstyle \frac{1}{2}} i 
\alpha_\Lambda \Delta_H^2 [9 \Pi''_H(p) + \Pi''_G(p)]} 
\end{align}
to this order. The polarization diagrams are given by standard loop-integral 
calculations \cite{Podolsky2011, Katan2015} as 
\begin{align}
\label{G}
\Pi_G(p) & = \frac{N-1}{8 \pi^2 c^3} \left[ 1 + \ln \left( \frac{\Lambda_0^2}
{p^2} \right) - i \pi\theta(p^2) \right] \! , \\
\label{H}
\Pi_H(p) & = \frac{1}{8 \pi^2 c^3} \left[ 1 + \ln \left( \frac{\Lambda_0^2}
{\Delta_H^2} \right) \right. \\ & \;\;\;\;\;\;\;\;\;\;\;\;\;\;\;\; \left.
 - i \pi \frac{\sqrt{p^2 - 4 \Delta_H^2}}{\sqrt{p^2}} \theta(p^2 - 4 \Delta_H^2) 
\right] \! ,
\end{align}
where again $p^2 = \omega^2 - {\bm p}^2$ and $\theta$ is the Heaviside theta 
function.

The spectral function for spin excitations is given by the imaginary part of 
the response function. To analyze the line width of the Higgs mode at its 
energy minimum, which occurs at spatial momentum $\bm p = \bm 0$ (relative 
to the ordering wave vector, ${\bf Q}$), we show in Fig.~\ref{chiV} the 
quantity $- \chi_{\sigma\sigma}''(\omega,\bm 0)$. The spectral function has a 
Lorentzian shape, whose full-width at half-maximum height gives a decay width
\begin{align}
\label{fidelity}
\notag \Gamma_H^v (|\delta g|) & = \frac{\alpha_\Lambda}{8\pi c^3} \Delta_H 
(|\delta g|) \\ & = \frac{\alpha_0 \Delta_H (|\delta g|)}{8 \pi c^3 \left[ 
1 + \frac{(N+8)}{8 \pi^2 c^3} \alpha_0 \ln (\sqrt{2} \Lambda_0/\Delta_H) 
\right]}.
\end{align}
The first equality corresponds exactly to the width deduced from the Fermi 
golden rule in Ref.~\cite{Kulik2011} and the second uses the form of the 
running coupling constant deduced in Eq.~(\ref{RunningAlpha}); we stress 
that $\Delta_H$ contains further intrinsic dependence on $\alpha_\Lambda$. 
Physically, the dominant peak in Fig.~\ref{chiV} corresponds to the process 
where a Higgs mode decays spontaneously into two Goldstone modes, given by 
$\Pi''_G (\omega,\bm 0)$, while process of decay into two Higgs modes, 
$\Pi''_H (\omega,\bm 0)$, has a threshold at $\omega = 2 \Delta_H$ and does 
not to contribute to the line width, $\Gamma_H^v$.

\begin{figure}[t]
\includegraphics[width=0.425\textwidth,clip]{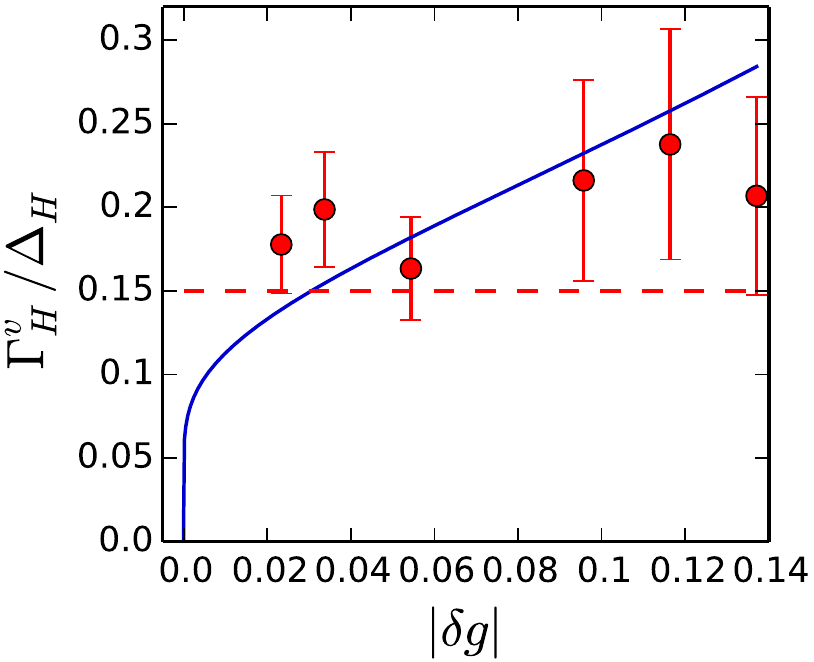}
\caption{Ratio $\Gamma_H^v/\Delta_H$ of the Higgs line width, as determined 
from the vector response function, to its gap, shown as a function of $|\delta 
g|$. The solid line is the QFT result obtained from Eq.~(\ref{fidelity}). The 
dashed line is the ratio extracted from QMC data by averaging over $|\delta 
g|$ and extrapolating in system size ($L \rightarrow \infty$) \cite{Qin2017}. 
The points are obtained from the QMC data for systems of sizes $L = 14$ and 16 
at the different values of $|\delta g|$ for which simulations were performed.}
\label{widthcomparison}
\end{figure}

Clearly the Higgs decay width in the vector channel is determined completely 
by the fundamental parameters of the QFT. We use the best-fit parameters 
[(\ref{bestfit1}) and (\ref{bestfit2})] for the double-cubic model to predict 
the Higgs line width (\ref{fidelity}) as a function of $|\delta g|$ and show 
the results as the solid line in Fig.~\ref{widthcomparison}. For the 
width-to-gap ratio, we find a function with approximately linear dependence 
in the range $0.04 < |\delta g| < 0.2$, but which falls sharply to zero once 
$|\delta g| < 0.02$. This latter behavior is dictated by the logarithmic terms 
in the running coupling constant and is in accord with the asymptotic freedom 
of the QFT at the QCP.

In Fig.~\ref{widthcomparison} we show also the width-to-gap data deduced 
from the QMC simulations of Ref.~\cite{Qin2017}. It is apparent immediately 
that the statistical errors in the numerical results are large on the scale 
of the changes in this quantity. Because data obtained for different system 
sizes, $L$, showed a spread significantly greater than the spread resulting 
from the different $|\delta g|$ values, the data were analyzed by averaging 
over $|\delta g|$ and extrapolating the results to large $L$. The resulting 
estimate of a constant ratio, $\Gamma_H^v/\Delta_H = 0.15$ (dashed line in 
Fig.~\ref{widthcomparison}), is equivalent to neglecting the logarithmic 
terms in Eq.~(\ref{fidelity}), and is consistent with experimental observations 
on TlCuCl$_3$ \cite{Ruegg2008,Merchant2014}. The quantitative analysis made 
possible by QFT demonstrates that observing the dominant logarithmic 
corrections to the width-to-gap ratio requires values of $|\delta g|$ not 
currently accessible to numerics or experiment. 

However, beyond the inaccessible regime $|\delta g| < 0.02$, it is possible 
to perform an alternative analysis of the QMC data informed by the QFT results. 
The data points with error bars in Fig.~\ref{widthcomparison} are obtained by 
considering the six $\delta g$ values individually. Instead of extrapolating 
in $L$, which would present very large errors, we retain only the two 
largest $L$ values ($L = 14$ and 16) and show their error-weighted average.
Despite the limitations of the numerical data, the matching trends of QFT and 
QMC illustrated in Fig.~\ref{widthcomparison} suggest that future QMC studies 
with only factor-2 improvements in the error bars in $|\delta g|$ could 
indeed demonstrate the logarithmic form of the Higgs decay width obtained 
from the vector response function.

\subsection{Scalar response function}

Turning now to the scalar response function, $\chi_{\varphi^2\varphi^2} (p) = 
\langle \vec\varphi^2(p) \, \vec\varphi^2(0) \rangle$, we use again the 
substitution $\vec\varphi = (\varphi_c + \sigma, \vec\pi)$ to effect a 
decomposition into Higgs and Goldstone components,
\begin{align}
\label{chiexpansion}
\notag \chi_{\varphi^2\varphi^2} (p) & = 4 \varphi_c^2 \chi_{\sigma\sigma} (p) + 4 
\varphi_c [\chi_{\sigma\pi^2} (p) + \chi_{\sigma\sigma^2} (p)] \\ & \;\;\;\; + 
\chi_{\sigma^2\sigma^2} (p) + 2 \chi_{\sigma^2\pi^2} (p) + \chi_{\pi^2\pi^2} (p).
\end{align}
Assisted by our results for the vector response (Sec.~\ref{Decay}A), we 
consider only the Higgs contributions to $\chi_{\varphi^2\varphi^2}$ at order 
$\alpha$; an alternative derivation may be found in Refs.~\cite{Podolsky2011, 
Katan2015}. We note first that 
\begin{align}
\notag \chi_{\pi^2\pi^2} (p) & = \langle \pi^2 (p) \pi^2 (0) \rangle = \Pi_G (p) \\
\notag \chi_{\sigma^2\sigma^2} (p) & = \langle \sigma^2 (p) \sigma^2 (0) \rangle = 
\Pi_H (p) 
\end{align}
are simply the Goldstone and Higgs polarization loops represented graphically 
on the bottom line of Fig.~\ref{Vector}, which are  given respectively by 
Eqs.~(\ref{G}) and (\ref{H}). 

For a first-order expansion of the other terms in $\chi_{\varphi^2\varphi^2}$, it 
is necessary to consider the form of coupling terms between the different 
fields allowed by the interaction, as specified in Eq.~(\ref{Lint}). In the 
case of the second term in Eq.~(\ref{chiexpansion}), we obtain 
\begin{align}
\notag 4 \varphi_c(\chi_{\sigma\pi^2} + \chi_{\sigma\sigma^2}) & = 4 \varphi_c 
(\langle \sigma \pi^2 \rangle + \langle \sigma\sigma^2 \rangle  + \langle 
\sigma [\alpha \varphi_c \sigma \pi^2]\pi^2 \rangle \\ \notag & \;\;\;\;\;\;\;
\;\;\;\;\;\;\;\;\; + \langle \sigma [\alpha \varphi_c \sigma \sigma^2]\sigma^2 
\rangle) \\ \notag & = 4 \alpha \varphi^2_c (\langle \sigma\sigma \rangle 
\langle \pi^2 \pi^2 \rangle  + 3 \langle \sigma\sigma \rangle \langle \sigma^2 
\sigma^2 \rangle) \\
 & = 4 \alpha \varphi^2_c \chi_{\sigma\sigma} (\Pi_G + 3\Pi_H).
\end{align}
Here the terms $\langle \sigma \pi^2 \rangle = \langle \sigma\sigma^2 \rangle
 = 0$ because there is no zeroth-order coupling of these fields. For the two 
terms in the second line, the insertions $[\alpha \varphi_c \sigma \pi^2]$ and 
$[\alpha \varphi_c \sigma \sigma^2]$ show the only terms in Eq.~(\ref{Lint})
coupling the fields at first order in the perturbative expansion. By the same 
reasoning, 
\begin{eqnarray}
\chi_{\sigma^2\sigma^2} & = & \langle \sigma^2 \sigma^2 \rangle + (\alpha 
\varphi_c)^2 \langle \sigma^2 [\sigma^2 \sigma \sigma \sigma^2] \sigma^2 \rangle 
\nonumber \\ & = & \Pi_H + 9 (\alpha \varphi_c)^2 \Pi_H \chi_{\sigma\sigma} \Pi_H, 
\\ \chi_{\pi^2\pi^2} & = & \langle \pi^2 \pi^2 \rangle + (\alpha\varphi_c)^2 
\langle \pi^2 [\pi^2 \sigma \sigma \pi^2] \pi^2 \rangle \nonumber 
\\ & = & \Pi_G + (\alpha \varphi_c)^2 \Pi_G \chi_{\sigma\sigma} \Pi_G, \\
2 \chi_{\sigma^2\pi^2} & = & 2 (\alpha \varphi_c)^2 \langle \sigma^2 [\sigma^2 
\sigma \sigma \pi^2] \pi^2 \rangle \nonumber \\ & = & 3(\alpha \varphi_c)^2 
\Pi_H \chi_{\sigma\sigma} \Pi_G,
\end{eqnarray}
and hence by summing all contributions in Eq.~(\ref{chiexpansion}) we obtain 
\begin{eqnarray}
\label{chiSpert}
\chi_{\varphi^2\varphi^2} & = & 4 \alpha \varphi^2_c \chi_{\sigma\sigma} [1 + \alpha 
(\Pi_G + 3\Pi_H)  \\ & & \;\; + {\textstyle \frac{1}{4}} \alpha^2 (\Pi_G^2 + 
9 \Pi_H^2 + 6 \Pi_G \Pi_H)] + \Pi_G + \Pi_H. \nonumber 
\end{eqnarray}

It is clear from the perturbative procedure that the divergent part of the 
scalar response function is linearly proportional to the vector response, 
$\chi_{\sigma\sigma}$ [Eq.~(\ref{spectralvector}) and Fig.~\ref{Vector}], and 
hence will share its pole structure. We note again that the terms $\Pi_G$ and 
$\Pi_H$ appearing in Eq.~(\ref{chiSpert}) have both real and imaginary parts, 
the first of which are responsible for the renormalization of the quantities 
$\varphi_c$ and $\Delta_H$, leading to the logarithmic corrections discussed 
in Sec.~\ref{QFToneloop}. Again we absorb these real parts into $\varphi_c$ 
and $\Delta_H$, showing only the imaginary parts, $\Pi_G''$ and $\Pi_H''$, 
which do not influence the renormalization. The final expression for the 
scalar response function is then 
\begin{align}
\label{chi}
\notag \chi_{\varphi^2\varphi^2} (p) & = \frac{2\Delta_H^2}{\alpha_\Lambda} 
\frac{ \{1 + \frac{1}{2} \alpha_\Lambda [\Pi''_G (p) + 3 \Pi''_H (p) ] \}^2}
{p^2 - \Delta_H^2 - \frac{i}{2} \alpha_\Lambda \Delta_H^2 \left[\Pi''_G(p)
 + 9\Pi''_H(p) \right]} \\ & \;\;\;\;\;\;\;\; + \Pi''_G (p) + \Pi''_H (p),
\end{align}
and the imaginary part of this quantity, which is the zone-center dimer-dimer 
spectral function of the double-cubic model, is shown as a function of the 
frequency $\omega$ at relative spatial wave vector ${\bm p} = {\bm 0}$. 

\begin{figure}[t]
\includegraphics[width=0.425\textwidth,clip]{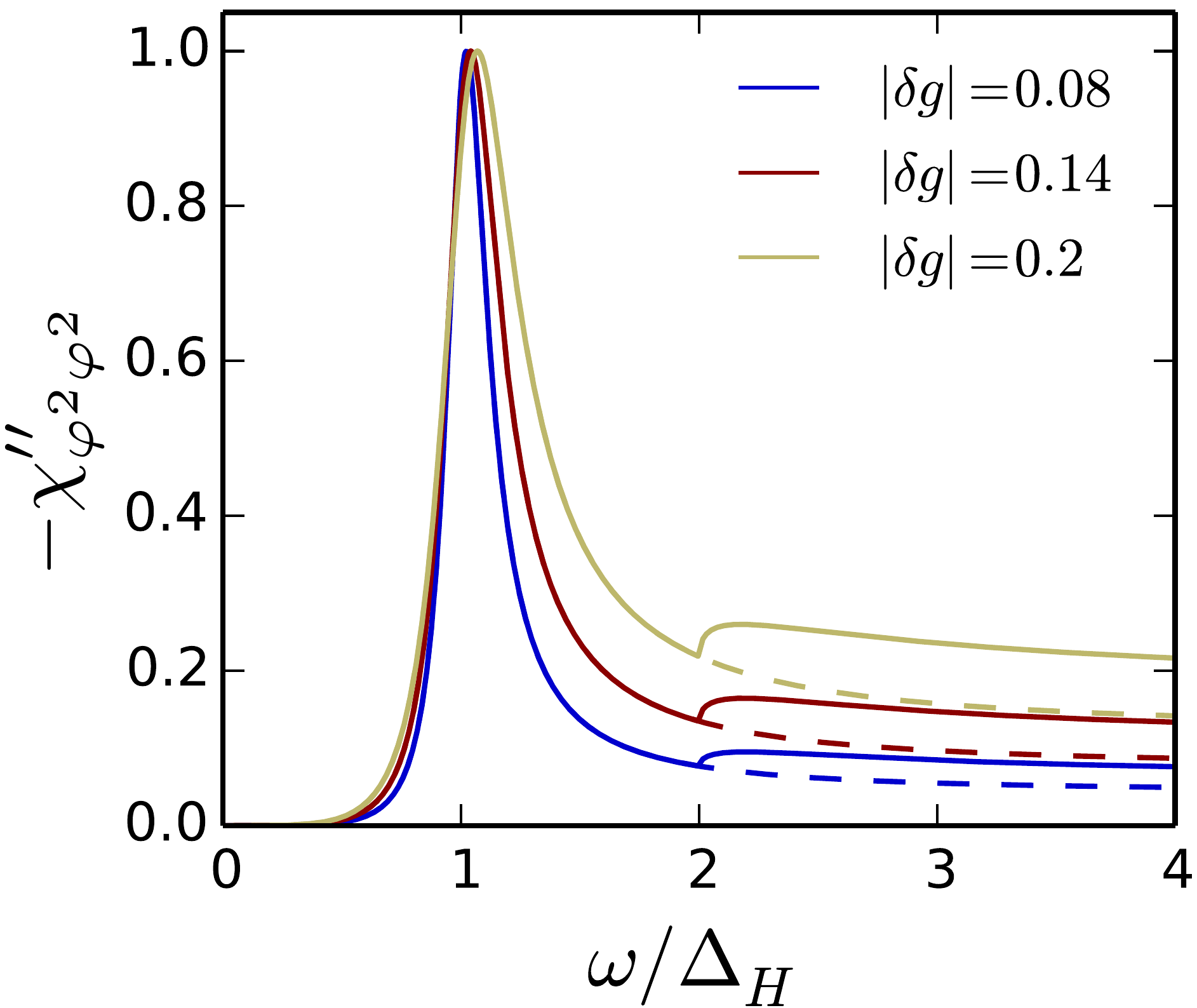}
\caption{Imaginary part of the scalar response function, $-\chi''_{\varphi^2\varphi^2} (\omega)$, shown as a function of $\omega/\Delta_H$ 
at $\bm p = \bm 0$ and normalized to its maximum value. The curves are 
evaluated from Eq.~(\ref{chi}) and correspond to different values, $|\delta 
g|$, of the coupling ratio relative to the QCP. Dashed lines show the results 
obtained from Eq.~(\ref{chi}) but neglecting the two-Higgs contribution 
[Eq.~(\ref{H})].}
\label{chiS}
\end{figure}

Several comments are in order concerning this result. First, the pole 
structure of the scalar response function is indeed identical to the 
vector response. The only difference to the spectral function is a prefactor 
arising from the imaginary part of the first term of Eq.~(\ref{chi}). Second, 
there are non-resonant pole contributions to $\chi_{\varphi^2\varphi^2} (\omega)$, 
which are contained in the lower line of Eq.~(\ref{chi}). In the limit of 
large four-momentum, $p^2 \gg \Delta_H^2$, these terms are dominant and the 
background scattering they contribute can be shown from Eqs.~(\ref{G}) and 
(\ref{H}) to have the asymptotic form
\begin{align}
\label{background}
\Pi''_G (p) + \Pi''_H (p^2 \gg \Delta_H^2) & \longrightarrow \frac{3}{8\pi}.
\end{align}
Setting $p = (\omega,\bm 0)$ in Eq.~(\ref{background}) accounts for the 
spectral weight of the high-$\omega$ tail in Fig.~\ref{chiS}.

Third, the phase of the prefactor and non-resonant pole terms contribute to a 
destructive interference in the emission channel of two low-energy Goldstone 
modes. This interference suppresses the imaginary part of the scalar response,  
resulting in the power-law form $\chi_{\varphi^2\varphi^2} (p) \propto p^4$ as 
$p \rightarrow 0$ in the present 3+1D problem, which is a statement of the 
Adler theorem. To show this explicitly, we reexpress the imaginary part of 
Eq.~(\ref{chi}) in the form, valid for $p^2 < 4 \Delta_H^2$,
\begin{align}
\label{adler}
\chi''_{\varphi^2\varphi^2} & = \frac{- p^4 \Pi''_G(p)}{(p^2 - \Delta_H^2)^2 + 
[\frac{1}{2} \alpha_\Lambda \Delta_H^2 \Pi''_G(p)]^2}.
\end{align}
Here we have neglected the $\Pi_{H}(p)$ term, which makes no contribution to 
the imaginary part for $p^2 < 4 \Delta_H^2$. The line shape of the scalar 
response function at ${\bm p} = {\bm 0}$, shown in Fig.~\ref{chiS}, is that 
of a Fano resonance, but with additional interference contributions that 
result in an $\omega^4$ form of the infrared tail \cite{Podolsky2011}. 
This asymmetric shape compares well with recent QMC results \cite{Qin2017, 
Lohofer2017}. However, our inclusion of the logarithmic corrections prevents 
any collapse of either the scalar or the vector response curves to a single 
`universal' form, as suggested by some of the QMC data. 

\begin{figure}[t]
\includegraphics[width=0.425\textwidth,clip]{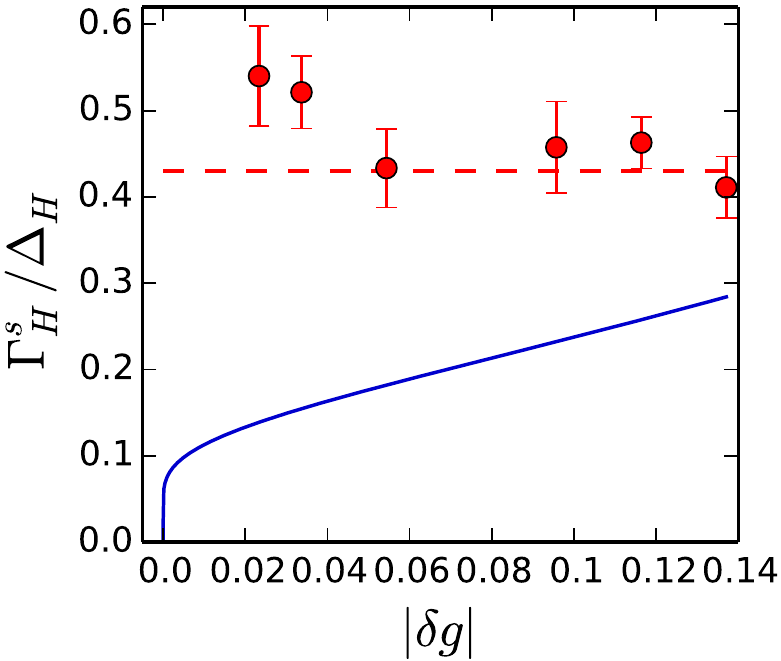}
\caption{Ratio $\Gamma_H^s/\Delta_H$ of the Higgs line width, as determined 
from the scalar response function, to its gap, shown as a function of $|\delta 
g|$. The solid line is the QFT result obtained from Eqs.~(\ref{chi}) and 
(\ref{fidelity}). The dashed line is the ratio extracted from QMC data by 
averaging over $|\delta g|$ and extrapolating in system size ($L \rightarrow 
\infty$) \cite{Qin2017}. The points are obtained from the QMC data for systems 
of sizes $L = 14$ and 16 at the different values of $|\delta g|$ for which 
simulations were performed.}
\label{Swidthcomparison}
\end{figure}

Finally, we comment on the Higgs decay width extracted from the scalar 
response function of Eq.~(\ref{chi}). While the asymmetric, non-Lorenzian 
shape of the dimer-dimer spectral function (Fig.~\ref{chiS}) prevents us 
from obtaining a direct analytic expression, to a good approximation the 
line width is still the value determined directly from the imaginary part 
of the denominator in Eq.~(\ref{chi}), which is identical to the result 
for the vector response (\ref{spectralvector}). Thus we obtain 
(\ref{fidelity}) $\Gamma^{s}_H \approx \Gamma^{v}_H = \alpha_\Lambda 
\Delta_H / 8\pi c^3$. 

In Fig.~\ref{Swidthcomparison} we plot the ratio $\Gamma_H^s/\Delta_H$, 
which for QFT is identical to the curve in Fig.~\ref{widthcomparison}, 
for comparison with the results obtained from the QMC simulations of 
Ref.~\cite{Qin2017}. Following the same procedure of averaging over 
$|\delta g|$ and extrapolating to large $L$ led to an anticipated constant 
ratio $\Gamma_H^s/\Delta_H = 0.43$, as shown by the dashed line. We show 
again the alternative analysis of retaining the individual $|\delta g|$ 
data and considering only the largest accessible values of $L$. In this 
case the QFT and QMC results differ very significantly, not only in magnitude 
but also in apparent functional form. Such a discrepancy cannot be ascribed 
solely to statistical errors in the QMC data and make clear that some 
systematic factors are also at work; one may speculate for example that 
the error bars on the imaginary-time QMC data have a particularly strong 
broadening effect in the SAC procedure for the scalar response function, 
in a way that does not affect the vector response. Further QMC and analytic 
continuation studies, including with simulated data obeying different error 
criteria, may be used to test such a hypothesis for different spectral 
functions. 

In summary, the extraction of the Higgs line width lies at the limits 
of current QMC data. Their accuracy is not yet sufficient to discern 
logarithmic corrections in this quantity from the vector response function  
(Fig.~\ref{widthcomparison}), while only the qualitative nature of the scalar 
response function is accessible (Fig.~\ref{chiS}). There has to date been no 
theoretical expectation with which to compare these results, and thus 
the present QFT analysis provides an essential quantitative benchmark. 
It is certainly desirable for future numerical studies to focus on the 
logarithmic dependence of the Higgs line width, which ultimately is expected 
because the theory becomes asymptotically free at the QCP. 

\section{Microscopic derivation of QFT parameters}\label{derivation}

The Lagrangian field theory (\ref{Lagrangian}) is a low-energy approximation 
to the full physics of the Hamiltonian (\ref{Hamiltonian}). Although we have 
shown that Eq.~(\ref{Lagrangian}) delivers an excellent description of the 
unbiased numerical data obtained by QMC simulations, the parameters we have 
used in making this comparison are fitted, and hence rank as phenomenological, 
i.e.~an explicit connection to the ``fundamental'' parameters, $J$ and $J'$ 
of the underlying spin model (\ref{Hamiltonian}) is lost. Here we employ a 
microscopic description, the bond-operator framework, to demonstrate the 
bridging of this gap between QFT and the spin Hamiltonian. Specifically, we 
will derive expressions for $\gamma$, $c$, and $g_c$ directly in terms of $J$ 
and $J'$ and provide an analytic justification for the linear relationship 
between $\varphi_c$ of QFT and $m_s$ of the spin Hamiltonian. However, this 
analytic treatment does not provide results for the arbitrary normalization 
points $\alpha_0$ and $\Lambda_0$ of the QFT. 

\subsection{Triplon gap, velocity, and the QCP}

The bond-operator representation \cite{Sachdev1990} is an identity for 
spin-$1/2$ operators that is particularly well adapted to the analysis of 
dimerized quantum magnets \cite{Gopalan1994}. When all the spins of the system 
reside on one dominant bond, as in Eq.~(\ref{Hamiltonian}) when $g \gg g_c$, 
it is logical to express the spin degrees of freedom as 
\begin{align}
\label{BOtransform}
{\bm S}^{l,r}_i & = {\textstyle \frac{1}{2}} (\pm s_i^{\dag}t_{i,\alpha} \pm 
t_{i,\alpha}^{\dag} s_i - i \epsilon_{\alpha,\beta,\gamma}t_{i,\beta}^\dag t_{i,\gamma})
\end{align}
where $s_i^{\dag}$ is an operator creating the singlet state of the two spins 
on bond $i$ and $t_{i,\alpha}^{\dag}$ creates one of the three triplet states. 
These singlet and triplet states have bosonic commutation relations, but 
from the nature of the underlying spin degrees of freedom are nevertheless 
mutually exclusive (i.e.~ they are hard-core bosons \cite{Sachdev1990}). 
When a system is strongly dimerized, its ground state may be treated as 
a condensate of bond singlets whose coherence is mediated by the hopping 
of (well gapped) triplet excitations, and hence it is an excellent
approximation to replace the operators $s_i^{\dag}$ and $s_i$ by their 
condensate expectation value, $\langle s \rangle = \bar{s}$. 

By applying the transformation of Eq.~(\ref{BOtransform}) to the QAF on the 
double-cubic lattice (\ref{Hamiltonian}) and performing standard Fourier and 
Bogoliubov transformations, we derive two mean-field equations whose 
self-consistent solution provides a quantitative description of the system 
for any coupling ratio, $g$. Full details of this procedure are provided in 
the Appendix. Although the two mean-field bond-operator parameters are in 
principle a function of $g$, we obtain a singlet condensation $\bar{s} = 
0.97$ for all values of $g$ in a broad region around the QCP. This includes 
the ordered phase, considered in the bond-operator formulation in 
Refs.~\cite{Matsumoto2002.2004} and \cite{Ruegg2008}, where the physical understanding 
of the magnetic state is small degree of antiferromagnetic order superposed 
on strongly fluctuating singlet correlations. For the present purposes, we 
focus on the bond-operator expression for the gap to triplon excitations in 
the quantum disordered phase, $\Delta_{\text{BO}} = \left( {\textstyle 
\frac{1}{4}} J' - \mu \right) [1 - 3d]^{1/2}$ [Eq.~(\ref{BOgap})], which we 
distinguish from $\Delta_t$ [Eq.~(\ref{gap})] obtained in QFT. Here $\mu$ 
is the other mean-field parameter, which corresponds to a triplon chemical 
potential, while $d$ is an average quantity depending linearly on $J$ and 
$\bar{s}^2$ as shown in the Appendix.

Having found two expressions for the triplon gap, one of which is given 
directly in terms of the fundamental parameters $J$ and $J'$, we can 
estimate the coefficient $\gamma$ in the QFT gap [Eq.~(\ref{gap})]. We  
equate the two gaps at the normalization point, $\Lambda_0$, to obtain 
the approximation
\begin{align}
\label{gammaValue}
\gamma^2 = \frac{\Delta^2_{\text{BO}}(\Lambda_0)}{|\delta g(\Lambda_0)|}.
\end{align}
Having chosen the normalization point $\Lambda_0 = 0.915 J$ on the basis 
of the criterion $\Delta_t (\Lambda_0) = \Lambda_0$, we find that $|\delta g 
(\Lambda_0)| \approx 0.056$ and thus obtain the estimate $\gamma = 3.88 J$. 
This compares rather well with the value $\gamma = 3.95 J$ obtained in 
Eq.~(\ref{bestfit1}), demonstrating that the phenomenological parameters 
required to fit the QMC data do indeed have a direct microscopic basis. 

The QCP in the bond-operator approach can be found by setting $\Delta_{\text{BO}}
 = 0$, which yields the value $g_c = 4.96$, in good agreement with the 
numerically exact result, $g_c = 4.83704$ \cite{Qin2015}. We also estimate 
the spin-wave velocity at the QCP from 
\begin{align}
c = \lim_{{\bm q} \rightarrow {\bm Q}} \frac{\Omega_{\bm q}(g_c)}{|{\bm q} - {\bm Q}|}
 = 2.28,
\end{align}
where $\Omega_{\bm q}(g)$ is the bond-operator triplon spectrum derived in 
the Appendix and $\bm Q$, the antiferromagnetic point in the Brillouin zone, 
is where the gap closes at $g_c$. Again we obtain good agreement with the 
QMC result, $c = 2.365$, demonstrating the quantitative accuracy of the 
bond-operator description. 

\subsection{Relationship of $m_s$ and $\varphi_c$}\label{UpsilonSec}

As noted in Sec.~\ref{Results}A, QFT cannot specify the staggered 
magnetization, $m_s$, directly, providing instead the order parameter, 
$\varphi_c$. To derive the relation between $m_s$ and $\varphi_c$, we 
consider the triplon bond operator, which we express as the vector 
$\vec{t}$, to find the constant of proportionality, $Z$, in the equation 
\begin{align}
\vec{\varphi} = {\textstyle \frac{1}{2}} Z^{-1} (\vec{t} \ ^\dag + \vec{t})
\end{align}
relating it to the vector field $\vec{\varphi}$. 
Working in real space, 
\begin{align}
\varphi(x) & = \sum_{\bm k} \frac{1}{\sqrt{2\Omega_{\bm k}}} \left[ \beta_{\bm k} 
e^{i kx} + \beta_{\bm k}^\dag e^{-i kx}\right] \! , \\
t(x) & = \sum_{\bm k} \left[ u_{\bm k} \beta_{\bm k} - v_{-{\bm k}} \beta_{-{\bm k}}^\dag 
\right] e^{i kx}, \\
& \approx \sum_{\bm k} \sqrt{\frac{A_{\bm k}}{2\Omega_{\bm k}}} \left[ \beta_{\bm k} 
e^{i kx} + \beta_{\bm k}^\dag e^{-i kx} \right] \! ,
\end{align}
where $\beta_k$ are the Bogoliubov operators diagonalizing the triplon 
Hamiltonian, $u_{{\bm k}}$ and $v_{{\bm k}}$ are the corresponding coefficients, 
defined in Eq.~(\ref{bogo}), and $A_{\bm k}$ (\ref{Bk}) are the diagonal 
components of the triplon matrix. In the vicinity of the QCP, the dominant 
contributions to the wave-vector sums are from low-energy excitations with 
${\bm k}$ of order ${\bm Q}$, allowing the approximation
\begin{align}
\label{Zfactor}
Z = \frac{1}{\sqrt{A_{\bm Q}}}.
\end{align}
The staggered magnetization of the QAF is 
\begin{align}
m_s^z & = \frac{1}{N} \langle S^{lz} - S^{rz} \rangle,
\end{align}
where $S^{zl,zr} = \sum_{i}^{N'} S^{zl,zr}_i$ with $N' = N/2$ the number 
of sites on each sublattice, whence 
\begin{align}
\notag m_s^z & = \frac{1}{2N} \left\langle \sum_{i\in A}^{N'} (s_i^{\dag} t_{i,z} + 
t_{i,z}^{\dag} s_i) + \sum_{i\in B}^{N'} (s_i^{\dag} t_{i,z} + t_{i,z}^{\dag} s_i) 
\right\rangle \\
& = {\textstyle \frac12} {\bar{s}} \langle t_z + t_z^\dag \rangle = \bar{s} Z 
\langle \varphi_z \rangle
\end{align}
and thus, because $\langle \varphi_z \rangle = \varphi_c$,
\begin{align}\label{UpsilonValue}
\Upsilon & = \frac{\varphi_c}{m_s^z} \; = \; \sqrt{\frac{1}{\bar{s}^2 A_{{\bm Q}}}}
\; = \; 0.62.
\end{align}
Once again we obtain a good microscopic account of the value $\Upsilon = 0.65$ 
deduced in Fig.~\ref{fitsgap}(b) by applying the QFT fitting framework to the 
QMC data. 

\section{Discussion}\label{discussion}


In summary, we have considered the critical properties of 3D quantum 
antiferromagnets as an example of a physical system at the upper critical 
dimension. The ability to obtain unbiased numerical data from QMC, of a 
precision high enough to verify multiplicative logarithmic corrections 
around the QCP in both static and dynamic observables, is an achievement 
at the frontier of current computational capabilities. By interpreting 
these data within the framework of an effective QFT, we obtain (i) unified 
physical insight into the connection between the static and dynamical 
properties of critical systems, (ii) a thorough test of perturbative O(3) 
QFT, and (iii) a valuable guide for the understanding of numerical and 
experimental studies probing quantum critical phenomena in a range of 
physical systems. 


At a pragmatic level, the present work offers a means for direct comparison 
between QMC and QFT. QMC data are obtained directly from the spin ($J$--$J'$) 
Hamiltonian (\ref{Hamiltonian}), whereas QFT results are derived in terms of 
the quasiparticles of a low-energy effective Lagrangian for long-wavelength 
fields (\ref{Lagrangian}). The excellent overall agreement demonstrates 
clearly the ability of the low-energy theory to capture all of the relevant 
physics in the vicinity of the QCP, and a quantitative description of the 
observables of the system allows the number of unknown parameters in the QFT 
to be reduced significantly. Once these fitting parameters are obtained, the 
QFT becomes predictive, which we demonstrate by calculating the vector and 
scalar response functions with an accuracy not currently achievable by QMC. 

It is well known from general QFT that the dimensionality and symmetry 
properties of a system determine the critical indices of its observables 
uniquely in the regime around the QCP. Previous numerical and experimental 
tests of universality have therefore focused on individual critical indices. 
In the present QFT analysis, we go beyond the asymptotic scaling behavior to 
provide a quantitative description of the observables and thus to investigate 
how they are connected. The crucial physical insight underlying unification of 
the thermodynamic and dynamic quantities is that the logarithmic corrections 
to their scaling specified in Eqs.~(\ref{observablesms})-(\ref{observablesTN}) 
may all be understood in terms of the running coupling constant 
(\ref{RunningAlpha}) between the quasiparticles of the QFT. 


Here we have focused primarily on the zero-temperature behavior of the 
system, as contained in the order parameter, gaps, and decay widths. Finite 
temperatures introduce thermal as well as quantum fluctuations and produce 
many exotic phenomena not present at zero temperature \cite{Sachdev1997, 
Sachdev1999, Sachdev2011, VojtaReview2003}. In particular, thermal 
fluctuations are responsible for the crossover into regions of the phase 
diagram marked as `quantum critical' in Fig.~\ref{phase}(a), where they 
interfere qualitatively with quantum effects, and are dominant in the region 
marked as `classical critical.' In these regimes, the observables of the 
system show different types of characteristic scaling behavior, to the point 
where the results of classical statistical mechanics are recovered. In this 
context it is crucial to remark that the finite-temperature behavior of the 
physical observables in QFT is completely determined by the results we 
present here (Sec.~\ref{QFToneloop} and Ref. \cite{ScammellFreedom2015}), 
i.e.~an analysis of finite-temperature properties would require no 
new fitting parameters. Because QMC is actually easier at finite temperatures, 
where no extrapolation is required in the corresponding system dimension, a 
quantitative investigation of static and dynamical properties across the 
full phase diagram by combining QFT and QMC is definitely feasible. 

Qualitatively, finite temperatures also generate additional scattering 
channels for quasiparticles, which are sometimes modelled as a heat bath. 
Among the physical implications of heat-bath scattering is the possibility 
that triplons in the disordered phase, which have infinite lifetimes at zero 
temperature, can acquire a substantial decay width. This situation has been 
investigated experimentally in TlCuCl$_3$ \cite{Merchant2014} and discussed 
analytically in Refs.~\cite{ScammellWidths2017, Fidrysiak2017} for the 
quantum antiferromagnet and Ref.~\cite{Nagao2016} for the Bose gas. A 
corresponding numerical (QMC) study of triplon decay at finite temperatures 
has yet to be performed.


A key additional direction for the extension of the present analysis is to 
include the effects of an applied magnetic field, which provides an explicit 
breaking of the spin symmetry. Early theoretical \cite{Nikuni2000} and 
experimental \cite{Ruegg2003} studies of the quantum antiferromagnet in 
the presence of a magnetic field investigated the phenomenology of magnon 
Bose-Einstein condensation, and suitably modified QFT descriptions have been 
used to discuss the associated critical scaling behavior \cite{Fisher1989, 
Sachdev1997, ScammellCritical2017}. Early QMC studies were also made of 
the magnon Bose-Einstein condensation scenario in 3D \cite{Wessel2001, 
Nohadani2004, Nohadani2005}, while some exotic theoretical predictions for 
quasi-1D systems \cite{Orignac2007} remain to be tested numerically. Once 
again, a QFT description of the critical observables can be obtained from 
the present work without the need for additional fitting parameters. Indeed, 
in a recent study of the 3D case, some of us \cite{ScammellCritical2017} 
predicted that two new critical indices emerge in the presence of an applied 
magnetic field and that logarithmic corrections are an important feature of 
the scaling behavior. To date there exists no related QMC analysis of a 
precision suitable for a comparative test.  

Finally, we anticipate that our results and techniques will serve as a 
helpful guide for future experimental and numerical studies of quantum 
critical phenomena. The reality of the situation is that research of the 
frontier of what is currently possible is always struggling for adequate 
data, by which is meant both enough data and sufficiently accurate data. 
The consequences of our results for numerical analysis include improved 
interpretation and understanding of critical regimes, the ability to relate 
datasets to reduce statistical errors, and qualitative guidance in previously 
unexplored but feasible directions. The additional consequences for experiment 
include the fact that all measurements in condensed matter and ultracold 
atomic condensates are made at finite temperature, and thus a systematic 
means of understanding the quantum limit is indispensible. 

\section*{Acknowledgments}

We are grateful to A. Sandvik for valuable contributions. HS, YK, and OPS 
were supported by the Australian Research Council under Grant No.~DP160103630. 
YQQ and ZYM were supported by the Ministry of Science and Technology of China 
under Grant No.~2016YFA0300502, the National Science Foundation of China under 
Grant Nos.~11421092 and 11574359, and the National Thousand-Young-Talents 
Program of China. 

\appendix*

\section{Bond-Operator Representation}

Here we provide details of the bond-operator technique and its application to 
the spin Hamiltonian of the 3D dimerized QAF (\ref{Hamiltonian}). As stated 
in Sec.~\ref{derivation}, the bond-operator representation of spins 
${\bm S}^{l,r}_i$ [Eq.~(\ref{BOtransform})] is particularly appropriate for a 
dimerized QAF. The most important point about the identity (\ref{BOtransform}) 
is that it must satisfy the SU(2) spin algebra,
\begin{align}
\notag [ S_{\alpha}^m, S_{\beta}^m] & = i \epsilon_{\alpha\beta\gamma} S_{\gamma}^m, 
\;\;\;\; [ S_{\alpha}^l, S_{\beta}^r] = 0,
\end{align}
which in fact sets two conditions on the bond operators $s_i^\dag$ and 
$t_{i,\alpha}^\dag$. One is that they must have bosonic commutation relations and 
the other that the space of physical states on any dimer bond constrains their 
total number to satisfy $s_i^\dag s_i + t_{i,\alpha}^\dag t_{i,\alpha} = 1$. However, 
satisfying this constraint on every dimer bond, $i$, leads to a problem that 
cannot be treated analytically and is extremely demanding numerically, but 
it has been shown \cite{Matsumoto2002.2004,Ruegg2008,Merchant2014} for the 3D QAF that 
satisfying the constraint only on average leads to quantitatively accurate 
results. This we effect using a Lagrange multiplier, $\mu$, that is the 
same on all sites \cite{Sachdev1990}. 

By applying the transformation of Eq.~(\ref{BOtransform}), the dimer-bond 
part of Hamiltonian (\ref{Hamiltonian}) becomes
\begin{align}
H_0 = J' \sum_i - {\textstyle \frac{3}{4}} s_i^{\dag} s_i + {\textstyle 
\frac{1}{4}} t_{i,\alpha}^{\dag} t_{i,\alpha} - \mu (s_i^\dag s_i + t_{i,\alpha}^\dag 
t_{i,\alpha} - 1).
\end{align}
The inter-dimer part contributes terms of higher order in the operators 
$s_i$ and $t_{i,\alpha}$ and, by retaining only those at quadratic order in 
$t_{i,\alpha}$, i.e.~by neglecting triplon interactions, we obtain 
\begin{align}
\label{H2}
H_2 & = {\textstyle \frac12} J \sum_{<i,j>} s_i^{\dag} s_j^\dag t_{i,\alpha} 
t_{j,\alpha} + s_i^{\dag} s_j t_{i,\alpha} t_{j,\alpha}^\dag + \text{H.c.}
\end{align}
This we treat in the approximation of complete Bose condensation of singlets, 
i.e.~we neglect singlet fluctuations and replace $s_i^\dag$ and $s_i$ by the 
constant $\bar{s}$.

The quadratic Hamiltonian $H_0 + H_2$ is expressed in reciprocal space using 
$t_{i,\alpha}^\dag = \frac{1}{\sqrt{N'}} \sum_{\bm k} t_{\bm k, \alpha} e^{-i\bm k \cdot 
\bm R_i}$, where $N' = N/2$ is the number of dimers, and diagonalized by a 
Bogoliubov transformations. The dynamical terms in the resulting Hamiltonian 
are
\begin{align}
\notag \bar{H}_2 & = \sum_{\bm k} A_{\bm k} t_{\bm k,\alpha}^\dag t_{\bm k,\alpha} + 
{\textstyle \frac{1}{2}} B_{\bm k} [t_{\bm k,\alpha}^\dag t_{-\bm k,\alpha}^\dag + 
\text{H.c.}] \\
& = \sum_{\bm k} \Omega_{\bm k} \beta_{\bm k, \alpha}^\dag \beta_{\bm k, \alpha},
\end{align}
where 
\begin{align}
\notag t_{\bm k,\alpha}^\dag & = u_{\bm k} \beta_{\bm k,\alpha}^\dag - v_{\bm k} 
\beta_{-\bm k,\alpha}, &&
\Omega_{\bm k} = \sqrt{A_{\bm k}^2 - B_{\bm k}^2} \\
\label{bogo}
u_{\bm k}^2, v_{\bm k}^2 & = \pm \frac{1}{2} + \frac{A_{\bm k}}{2\Omega_{\bm k}}, & &
u_{\bm k} v_{\bm k} = \frac{B_{\bm k}}{2\Omega_{\bm k}}.
\end{align}
The coefficients $A_{\bm k}$ and $B_{\bm k}$ depend on the lattice geometry and 
for the double-cubic model are 
\begin{align}
\notag A_{\bm k} & = {\textstyle \frac{1}{4}} J' - \mu + J\bar{s}^2 \, [\cos 
k_x + \cos k_y + \cos k_z] \\
\label{Bk}
B_{\bm k} & = J \bar{s}^2 \, [\cos k_x + \cos k_y + \cos k_z].
\end{align}

To obtain an expression for the triplon spectrum and hence the gap, it is 
necessary to deduce the mean-field parameters $\mu$ and $\bar{s}$, which  
are obtained from the saddle-point conditions
\begin{align}
\left\langle \frac{\partial H_{MF}}{\partial\mu} \right\rangle & = 0, 
& & \left\langle \frac{\partial H_{MF}}{\partial\bar{s}} \right\rangle = 0, 
\end{align}
in which $H_{MF} = \bar{H}_0 + \bar{H}_2$ denotes both the constant and 
dynamical parts of the quadratic mean-field Hamiltonian. It is convenient 
\cite{Gopalan1994} to introduce the dimensionless parameter 
\begin{align}
\label{aux}
d = \frac{2J\bar{s}^2}{{\textstyle \frac{1}{4}} J' - \mu},
\end{align}
in terms of which the self-consistent mean-field equations are 
\begin{align}
\notag \bar{s}^2 & = \frac{5}{2} - \frac{3}{2N'} \sum_{\bm k} 
\frac{1 + d\gamma_{\bm k}}{\sqrt{1 + 2 d \gamma_{\bm k}}}, \\
\mu & = - \frac{3J'}{4} + \frac{3J}{N'}\sum_{\bm k} 
\frac{\gamma_{\bm k}}{\sqrt{1 + 2 d \gamma_{\bm k}}}, 
\label{meanfield}
\end{align}
with 
\begin{align}
\notag d & = \frac{J}{J'} \left( 5 - \frac{3}{N'}\sum_{\bm k} 
\frac{1}{\sqrt{1 + 2 d \gamma_{\bm k}}} \right) \! , \\ \notag
\gamma_{\bm k} & = {\textstyle \frac{1}{2}} [\cos k_x + \cos k_y + \cos k_z].
\end{align}
The triplon spectrum may now be expressed as 
\begin{align}
\notag \Omega_{\bm k} & = \left( {\textstyle \frac{1}{4}} J' - \mu \right) 
[1 + 2 d \gamma_{\bm k}]^{1/2} 
\end{align}
and the gap as 
\begin{align}
\label{BOgap}
\Delta_{\text{BO}} & = \left( {\textstyle \frac{1}{4}} J' - \mu \right) [1 - 
3 d]^{1/2}.
\end{align}
This expression for $\Delta_{\text{BO}}$ was used to evaluate $\gamma$ in 
Eq.~(\ref{gammaValue}) and to derive the bond-operator value of the QCP, 
$J'/J = g_c = 4.96$; at values of $g$ around $g_c$, we obtain the result 
$\bar{s} = 0.97$, which was used in Eq.~(\ref{UpsilonValue}) to evaluate 
$\Upsilon$.


%

\end{document}